\documentclass[a4paper,12pt]{article}
\pdfoutput=1
\usepackage{jheppub}
\usepackage{graphicx}
\usepackage{placeins} 
\usepackage{siunitx}
\usepackage{hyperref}
\usepackage{amsmath}
\usepackage{ulem}

\title{Associated production of a $W$-boson and a charm meson at NNLO in QCD}
\author[a]{Terry Generet,}
\author[b]{Rene Poncelet,}
\author[c,d]{and Miha Muškinja}
\affiliation[a]{Cavendish Laboratory, University of Cambridge, Cambridge CB3 0HE, United Kingdom}
\affiliation[b]{Institute of Nuclear Physics, ul. Radzikowskiego 152, 31--342 Krakow, Poland}
\affiliation[c]{Faculty of Mathematics and Physics, University of Ljubljana, Jadranska ulica 19, 1000 Ljubljana, Slovenia}
\affiliation[d]{Jožef Stefan Institute, Jamova cesta 39, 1000 Ljubljana, Slovenia}

\emailAdd{generet@hep.phy.cam.ac.uk}
\emailAdd{rene.poncelet@ifj.edu.pl}
\emailAdd{miha.muskinja@fmf.uni-lj.si}

\abstract{
The production of heavy-flavor hadrons in association with a vector boson in proton-proton collisions is a powerful probe for studying Quantum Chromodynamics and the content of protons. In this article, we provide, for the first time, differential predictions through NNLO for the production of $W^{\pm}D^{(*)\mp}$ final states. The results are compared to recent ATLAS measurements, and the sensitivity to the strange content of the proton is investigated by PDF profiling. The results are found to be promising for including $W^{\pm}D^{(*)\mp}$ measurements in proton-proton collisions in PDF fits.
}


\preprint{Cavendish-HEP-25/04, IFJPAN-IV-2025-21}

\begin{document}

\maketitle

\section{Introduction}
The universal collinear (or integrated) quark and gluon parton distribution functions (PDFs) of the proton are essential inputs to phenomenological studies at colliders such as the Large Hadron Collider (LHC).
While attempts are made to derive them from first principles \cite{Lin:2017snn}, the most accurate and precise determinations are obtained from fits to data \cite{Alekhin:2017kpj, Hou:2019efy, Bailey:2020ooq, NNPDF:2021njg, PDF4LHCWorkingGroup:2022cjn}.
In particular, deep-inelastic scattering (DIS) processes, measured for example at HERA \cite{H1:2015ubc}, allow a direct measurement of structure functions and hence PDFs.
The available precision and dynamical range of the HERA data, however, make it necessary to include also data from the LHC itself to achieve the best possible precision.
The gluon and valence quarks ($d$ and $u$) are the best-determined PDFs, and modern fits achieve a few percent uncertainty in the most important phase space regions, while second (and third) family quark PDFs are much less constrained.

It is possible to gain sensitivity to individual quark PDFs in hadronic collisions by studying flavored final states, such as heavy-flavored hadrons or flavor-tagged jets.
The production of an electroweak vector boson ($\gamma$, $W^{\pm}$ or $Z$) in association with a heavy flavor quark-jet or hadron is a potent probe for individual quark PDFs due to the flavor structure of the hard-scattering process, see for example Figure \ref{fig:feyn}.

The production of $W^{\pm}$-boson in association with a charmed meson is of particular interest for the strange quark PDF \cite{Baur:1993zd, ATLAS:2014jkm, CMS:2018dxg, CMS:2021oxn, ATLAS:2023ibp, CMS:2023aim}.
Besides the role of the strange-quark PDF in the overall uncertainty budget in predictions for collider observables, the strange-anti-strange quark PDF asymmetry \cite{Catani:2004nc} is of particular interest because its observation would be a brilliant success of perturbative QCD predictivity.
Unfortunately, current precision does not yet permit a conclusive statement on this matter, and additional constraints from experimental data are needed.

There are two ways to view the production process of a $W^{\pm}$-boson plus charm flavor.
One possibility is to consider the formation of a jet arising from charm quarks that are produced with high energy.
This process has been computed through NNLO QCD plus NLO EW corrections \cite{Stirling:2012vh, Czakon:2020coa, Czakon:2022khx} and is available in all major Monte Carlo frameworks \cite{Alwall:2014hca, Bevilacqua:2021ovq, Sherpa:2024mfk, Bellm:2015jjp} at NLO, including parton shower simulations.
However, the $W^{\pm}$ plus a charm flavor tagged jet process requires careful treatment of the jet flavor definition \cite{Banfi:2006hf, Caletti:2022hnc, Caletti:2022glq, Czakon:2022wam, Gauld:2022lem, Caola:2023wpj, Behring:2025ilo} in the theoretical predictions.
In particular, beyond NLO QCD, flavor-sensitive jet algorithms are needed.
While recent investigations have solidified the theoretical understanding of the phenomenological impact of these modified jet algorithms, their implementation in experiments remains a subject of current research.
Unfolding procedures to translate between parton- and particle-level cross sections to interpret the data in terms of a PDF fit introduce additional complications and uncertainties.

The production of a vector boson in association with an identified charm hadron is an alternative approach to studying the same process, which does not rely on a jet definition.
From a theoretical perspective, this process can be described using collinear factorization.
In this case, the perturbative evolution of highly energetic partons is described through collinear DGLAP evolution \cite{Gribov:1972ri,Altarelli:1977zs,Dokshitzer:1977sg} and the non-perturbative transition of partons to hadrons through non-perturbative fragmentation functions \cite{Berman:1971xz}.
Such predictions have been performed through NNLO QCD for processes at lepton-lepton \cite{Rijken:1996vr,Rijken:1996ns,Rijken:1996npa,Mitov:2006wy,Blumlein:2006rr}, lepton-hadron \cite{Goyal:2023zdi,Bonino:2024qbh,Bonino:2024wgg,Goyal:2024tmo,Ahmed:2024owh,Goyal:2024emo, Bonino:2025tnf, Bonino:2025qta, Goyal:2025bzf,Bonino:2025bqa,Goyal:2025qyu}, and hadron-hadron colliders \cite{Czakon:2021ohs, Czakon:2022pyz,Bonino:2024adk, Czakon:2024tjr, Czakon:2025yti}.

In this work, we specifically study the phenomenology of $W^{\pm}$-bosons produced in association with $D^{(*)\mp}$-mesons and the implications for future PDF fits.
To that end, we perform, for the first time, NNLO QCD predictions for this process within the framework developed in Ref.~\cite{Czakon:2021ohs, Czakon:2025yti} and compare the results to recent ATLAS data \cite{ATLAS:2023ibp}.
The measurement was performed differentially in rapidity bins of the charged lepton originating from $W$-boson decays, which was found to be one of the most sensitive experimentally accessible observables for PDF determination in the $W^{\pm}D^{(*)\mp}$ process.
Although we do not perform a PDF fit in this work, we investigate the sensitivity to the strange quark content of the proton by PDF profiling.

PDFs depend on the perturbative order used in the DGLAP evolution as well as the partonic cross sections.
In light of the current work, it is essential to note that all modern PDFs are determined at least through NNLO QCD, making these results crucial for the consistent incorporation of $W^{\pm}D^{(*)\mp}$ data in such fits.
Beyond NNLO QCD, approximate N3LO PDFs have been obtained using N3LO DGLAP and NNLO QCD predictions with estimates for the missing higher-order terms \cite{McGowan:2022nag, NNPDF:2024nan}.

The article is organized as follows. In Section \ref{sec:setup}, we describe the computational setup, including all required PDF and FF inputs. We follow up with a discussion of LHC phenomenology through NNLO QCD as well as a comparison to ATLAS data in Section \ref{sec:pheno}. In Section \ref{sec:profiling}, we estimate the potential impact of using these observables in a PDF with the help of PDF profiling. Finally, we conclude in Section \ref{sec:conclusion}.

\section{Calculational setup}\label{sec:setup}

The computations presented in this work are performed within the four-dimensional sector-improved residue subtraction scheme \cite{Czakon:2010td, Czakon:2014oma, Czakon:2019tmo}.
The original scheme has been extended to allow for fragmentation processes in Ref.~\cite{Czakon:2021ohs, Czakon:2025yti}.
The subtraction is implemented in the well-tested \textsc{Stripper} C++ framework, which has been used for a range of similar computations \cite{Czakon:2021ohs, Czakon:2022pyz, Czakon:2024tjr, Czakon:2025yti}.
Besides the subtraction scheme we employ tree-level matrix elements from the \textsc{Avh} library \cite{Bury:2015dla}, one-loop matrix elements from \textsc{OpenLoops} \cite{Cascioli:2011va, Buccioni:2017yxi, Buccioni:2019sur}, and an implementation of the two-loop matrix elements for the $0 \to q\bar{q} W(\to \ell\nu)g$ processes \cite{Gehrmann:2011ab} that uses \textsc{Ginac} \cite{Bauer:2000cp, Vollinga:2004sn} to evaluate the appearing generalized polylogarithms.

\begin{figure}
    \centering
    \includegraphics[width=0.4\linewidth]{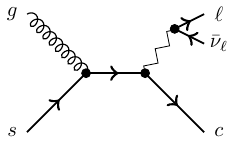}\hspace{0.1\linewidth}%
    \includegraphics[width=0.4\linewidth]{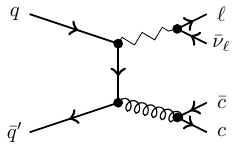}
    \caption{Example Feynman diagrams for the leading partonic contributions for $W$ plus charm production.}
    \label{fig:feyn}
\end{figure}

We set the electroweak parameters in the $G_{\mu}$ scheme and use the following numerical values:
\begin{itemize}
    \item Masses and widths: $m_W = \SI{80.379}{\giga\electronvolt}$, $m_Z = \SI{91.1876}{\giga\electronvolt}$, $\Gamma_W = \SI{2.085}{\giga\electronvolt}$, $\Gamma_Z = \SI{2.4952}{\giga\electronvolt}$. The Higgs mass and width parameters do not contribute.
    \item Coupling: $G_\mu = \SI{1.1663787e-5}{\giga\electronvolt^{-2}}$,
    \item We use the following CKM matrix (based on PDG \cite{ParticleDataGroup:2020ssz} values):
       \begin{align}
       V = \left(\begin{array}{ccc} 0.97401 & 0.22636 & 0. \\ 0.22650  & 0.9732 & 0. \\ 0.00361 & 0.0405 & 1.
       \end{array}\right)
       \end{align}
\end{itemize}

The (differential) cross sections depend on three unphysical scales: the renormalization scale $\mu_R$, the factorization scale $\mu_F$, and the fragmentation scale $\mu_{Fr}$, which we take all to be equal to
\begin{align}
  \mu_0 = \sqrt{m_W^2+p_T(W)^2}\;.
\end{align}
To give an estimate for missing higher-order uncertainties, we perform a 15-point scale variation around the central choice by varying each scale up and down by a factor of two, omitting extreme variations where any two scales differ by a factor of four. The scale uncertainties correspond to the envelope of the resulting 15 variations.

For the phenomenology studies in the Section \ref{sec:pheno} we employ the \textsc{MSHT20nnlo\-\_as118} PDF set~\cite{Bailey:2020ooq} (all perturbative orders are computed with the same NNLO set) as provided by the LHAPDF library \cite{Buckley:2014ana}, which we also use to evaluate $\alpha_s(\mu)$. For the profiling in Section \ref{sec:profiling} we additionally consider the \textsc{ATLASpdf21\_T1}~\cite{ATLAS:2021vod}. The motivation is that \SI{7}{\tera\electronvolt} $W$+charm data have already been used for the \textsc{MSHT20nnlo\-\_as118} set, whereas such data have not been used for the \textsc{ATLASpdf21\_T1} PDF set. Thus, these two PDFs represent different scenarios to study the additional constraining power from the \SI{13}{\tera\electronvolt} $W^{\pm}D^{(*)\mp}$ data set.

To suppress the contributions from final state $g \to c\bar{c}$ splittings (e.g. Figure~\ref{fig:feyn} right), all cross sections are defined as the difference between the \textit{opposite-sign} (OS) and \textit{same-sign} (SS) charge:
\begin{align}
\sigma^{\text{OS}-\text{SS}}(pp \to W^+D) &= \sigma(pp \to W^+D^-) - \sigma(pp \to W^+D^+)\;,\\
\sigma^{\text{OS}-\text{SS}}(pp \to W^-D) &= \sigma(pp \to W^-D^+) - \sigma(pp \to W^-D^-)\;.
\end{align}
The $D^{*\pm}$ cross sections are defined similarly.

We use the SKM18 \cite{Soleymaninia:2017xhc} FF set for $D^{*{\pm}}$ and SMSKA19 \cite{Salajegheh:2019nea} FF set for $D^{\pm}$.
For both sets, only the central fits are available without any variation that would allow consistent uncertainty estimates\footnote{In both cases, uncertainty estimates were obtained in the corresponding papers, but the complete uncertainty information was not published. We contacted the authors requesting this information, but, contrary to what is written in Ref.~\cite{Salajegheh:2019nea}, it is not available upon request.}.
To provide, nevertheless, an estimate of the sensitivity to the FF, we also use the AKSRV17 \cite{Anderle:2017cgl} FF set as a variation for $D^{*\pm}$ and use the difference between AKSRV17 and SKM18 as the FF uncertainty estimate.
For the $D^{\pm}$, we assume that the relative uncertainties are the same.
The OS-SS cross sections can be implemented on the level of the FFs by taking the difference between the $D^+$ and $D^-$ FFs for each partonic flavor.
To that end we obtain OS-SS FFs from the aforementioned FF sets and generate corresponding LHAPDF grids, using the APFEL library \cite{Bertone:2013vaa} to perform NNLL DGLAP evolution. We directly use these OS-SS FFs in the computation of the cross sections presented in this work.

All FF fits used here effectively treat charm fragmentation in full analogy to the fragmentation of light hadrons, using a separate empirical model for each parton type. Alternatively, one could use the perturbative fragmentation function formalism \cite{Mele:1990yq,Mele:1990cw} to perturbatively relate the fragmentation functions for different partons to each other. However, it was pointed out in Ref.~\cite{Bonino:2023icn} that the current understanding of non-perturbative effects plaguing that approach presently limit its theoretical accuracy for charm fragmentation. For this reason, we choose not to use the perturbative fragmentation function formalism for this study, using instead FF sets constrained only by data.

Consistent with our PDFs, we are using a massless charm quark in our computations. We investigated the impact of a finite charm quark and meson mass through NLO QCD and find it to be small, approximately $ 1\%$, and flat with respect to the rapidity of the charged lepton. Consequently, in the normalized lepton rapidity distributions used for the PDF profiling, these mass effects drop out. In our study, this will be the only case where a precision beyond 1\% is of phenomenological relevance.
For the transverse momentum distribution, the mass effect is about 1\% at low transverse momentum and further reduced in the tail of the distribution.
Finally, we studied the impact of $b\to B\to D$ decay chains, again at the OS-SS level, and they were found to be well below a permille and also mostly flat with respect to rapidity, so we ignore them for the present work.

\section{LHC Phenomenology}\label{sec:pheno}

We study the impact of higher-order corrections at the LHC with a center-of-mass energy of \SI{13}{\tera\electronvolt}. We focus here on a phase space definition which is consistent with the recent ATLAS measurement of $W^{\pm}D^{(*)\mp}$ production \cite{ATLAS:2023ibp}. To that end, we define the fiducial phase space by the following conditions:
\begin{itemize}
    \item at least one charged lepton with $\lvert\eta(\ell)\rvert<2.5$ and $p_T(\ell)>\SI{30}{\giga\electronvolt}$,
    \item at least one $D$-meson with $\lvert\eta(D)\rvert<2.2$ and $p_T(D)>\SI{8}{\giga\electronvolt}$.
\end{itemize}

\begin{figure}[t]
\includegraphics[width=0.50\textwidth]{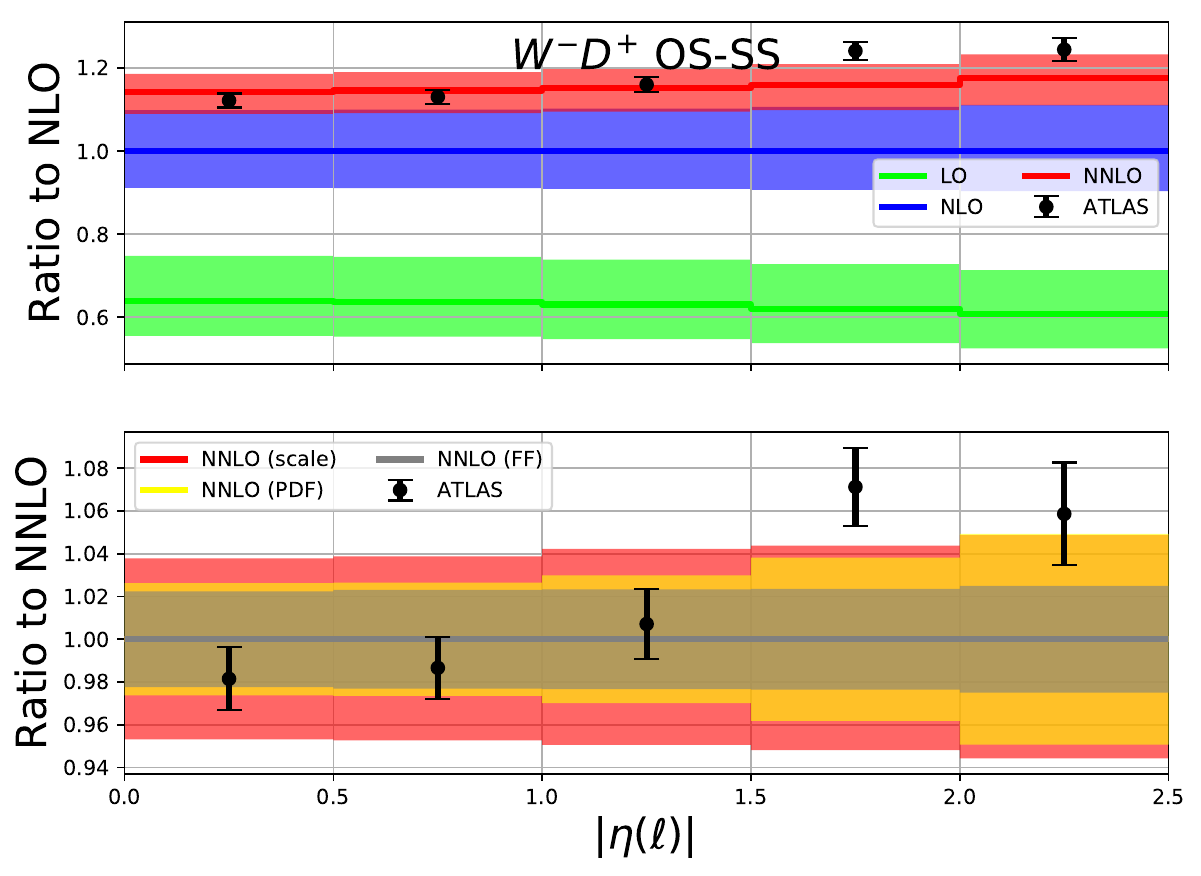}%
\includegraphics[width=0.50\textwidth]{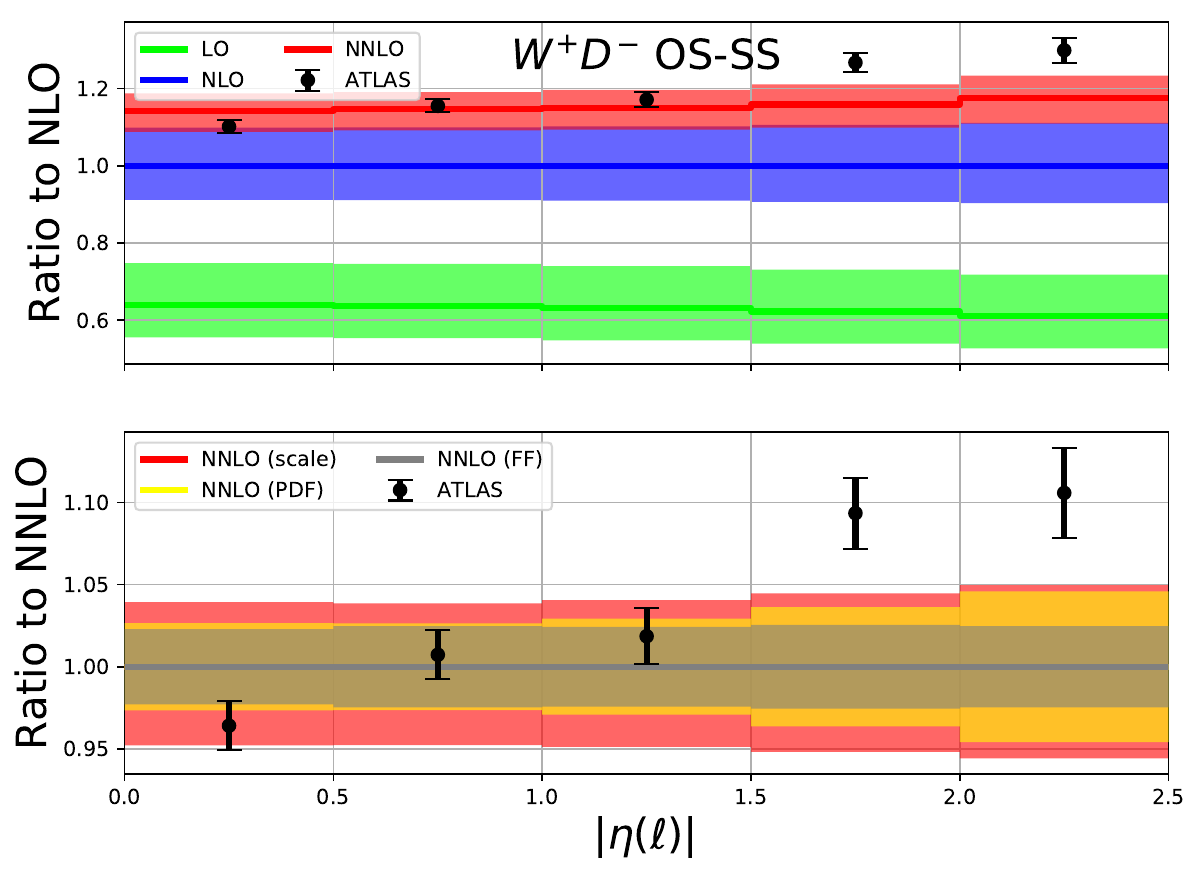}
\includegraphics[width=0.50\textwidth]{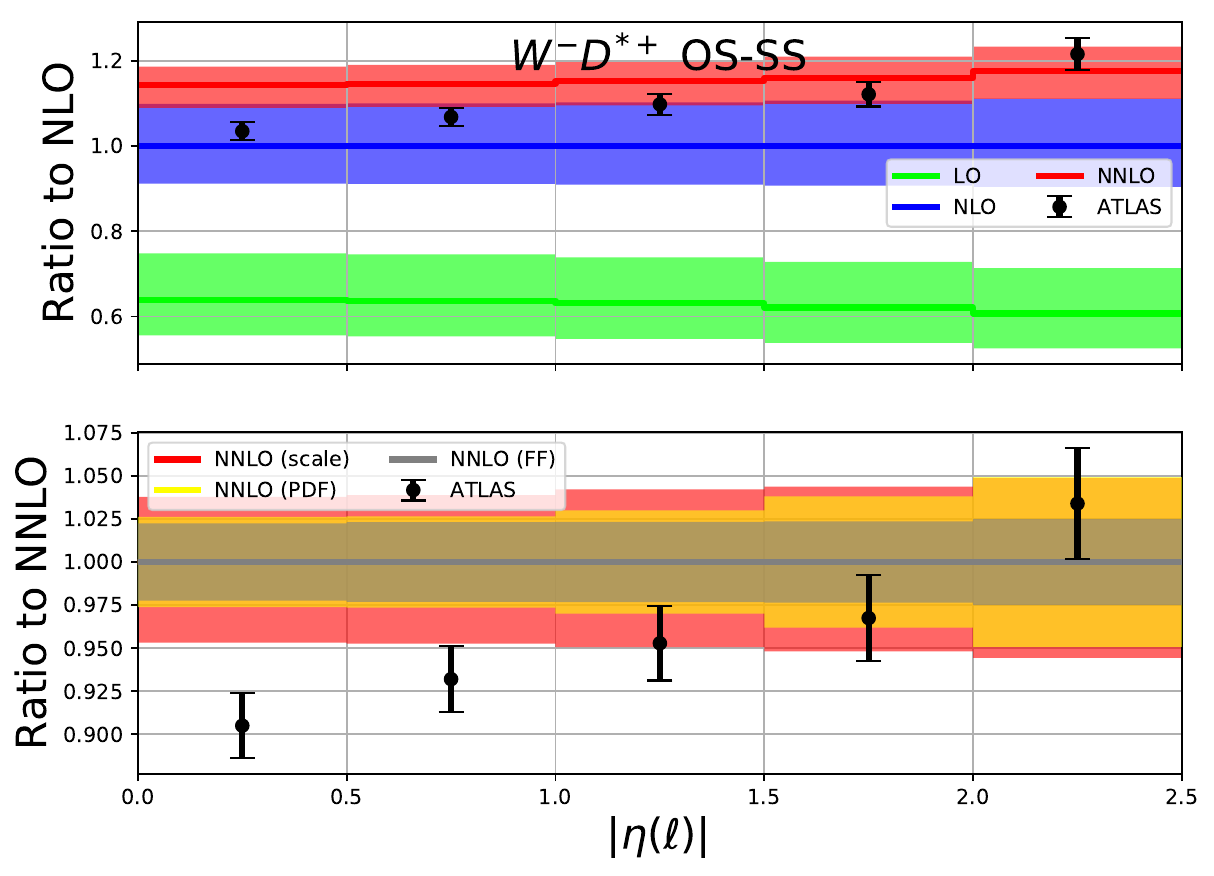}%
\includegraphics[width=0.50\textwidth]{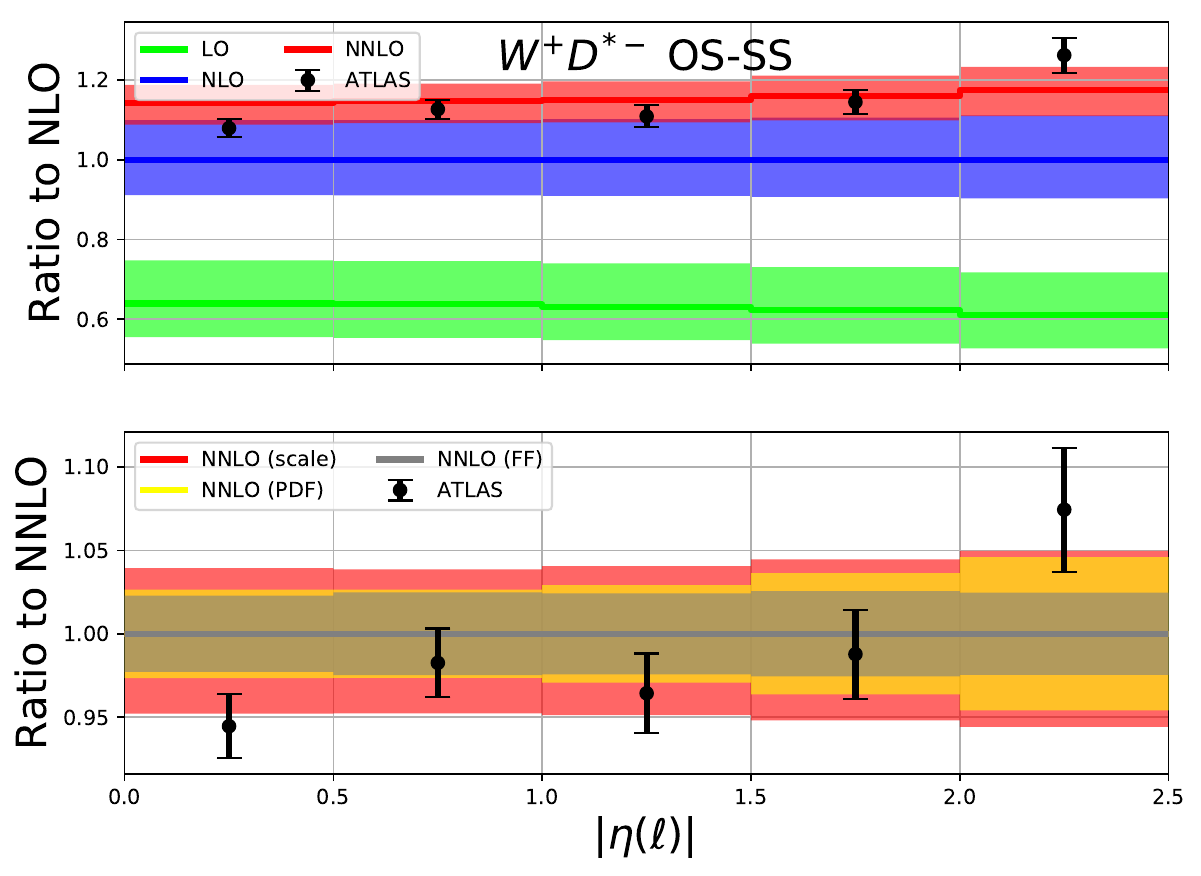}
\caption{Lepton pseudo-rapidity distributions for $W^-D^+$ (top left), $W^+D^-$ (top right), $W^-D^{*+}$ (bottom left) and $W^+D^{*-}$ (bottom right). The top panel of each plot shows the LO (green), NLO (blue) and NNLO (red) theory predictions, with coloured bands indicating the scale uncertainties. The black data points show the ATLAS data \cite{ATLAS:2023ibp}. The bottom panels show the NNLO predictions with three different sources of theory uncertainty: scale uncertainties (red), PDF errors (yellow), and FF errors (grey).}
\label{fig:rapidity}
\end{figure}

We begin our discussion with the absolute rapidity distribution, which is shown in Figure \ref{fig:rapidity}.
The figure is organized in four groups of plots, corresponding to each of the four final states $W^-D^+$, $W^+D^-$, $W^-D^{* +}$, and $W^+D^{*-}$ respectively.
Each set shows the theory predictions for the differential cross sections at LO, NLO, and NNLO QCD, normalized to the NLO QCD predictions in the top panel, along with an uncertainty breakdown at NNLO QCD in the lower panel.
Additionally, ATLAS data are also shown together with their respective experimental uncertainties.
We observe that the perturbative series converges well, reducing the theory uncertainties from $\pm10\%$ to about $\pm5\%$ when going from NLO to NNLO QCD for all charm hadron species.
The PDF uncertainties are of comparable size, in particular in the forward region, indicating the sensitivity of this observable to the proton content.
The estimated FF uncertainties are slightly smaller and completely flat, as the FFs essentially only enter as a normalization factor due to the negligible dependence of the fragmentation kinematics on rapidity.
The description of the data is improved by going from NLO to NNLO QCD, not only in normalization but also in shape.
The $D^{*+}$ case (the lower left plot) is the exception as NNLO QCD overshoots the measured cross section by about $5-10\%$.
However, in light of the size of the FF uncertainties and since the FF uncertainties for $D$ and $D^*$ are independent, it is possible for a more precise FF fit to obtain $D^*$ FFs that yield agreement with the data, without changing the predictions for $W^\pm D$, which already show reasonable agreement with the data.
The higher-order QCD predictions change the shape of the theory predictions mildly, improving towards the trends visible in the data; however, a shape difference remains.
This observation can be reconciled with the PDF sensitivity and interpreted as a potential pull on the PDFs if these observables are used in their extraction.

\begin{figure}[t]
\includegraphics[width=0.50\textwidth]{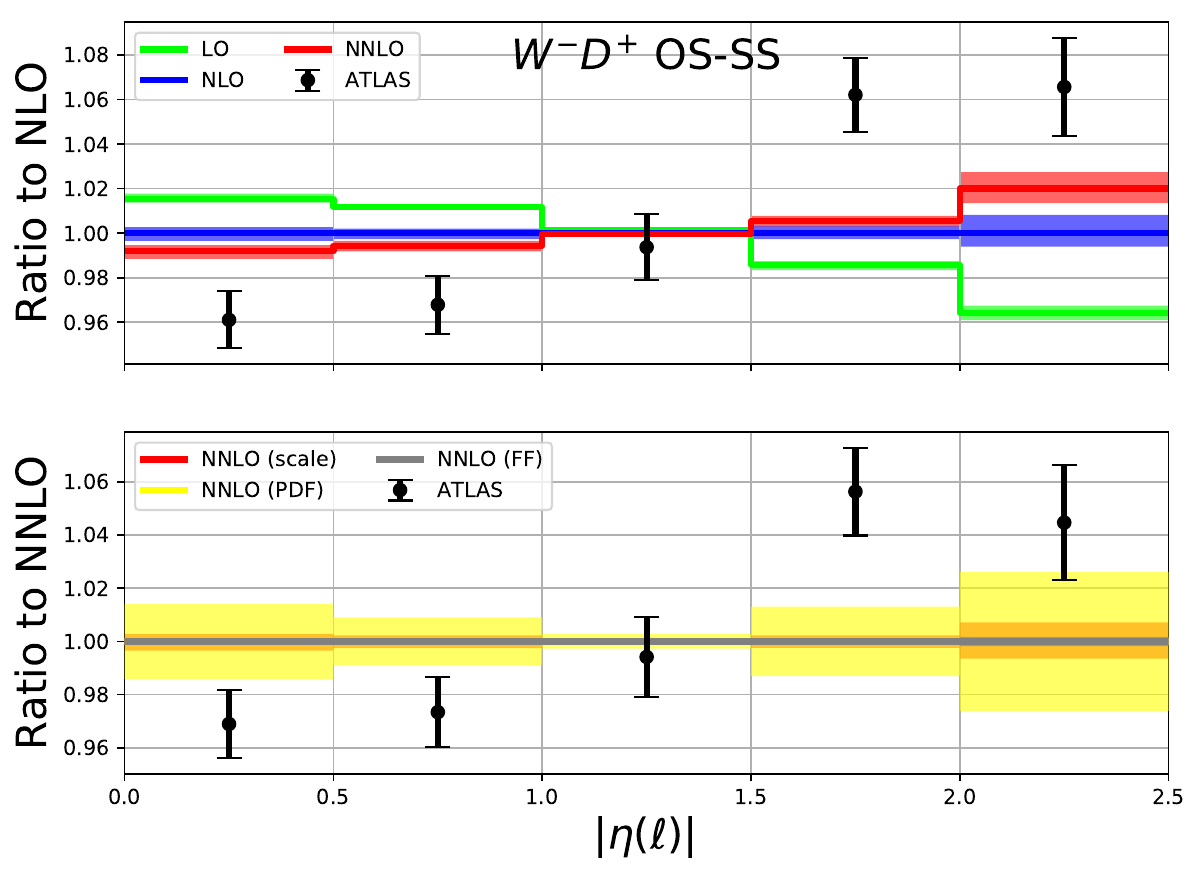}%
\includegraphics[width=0.50\textwidth]{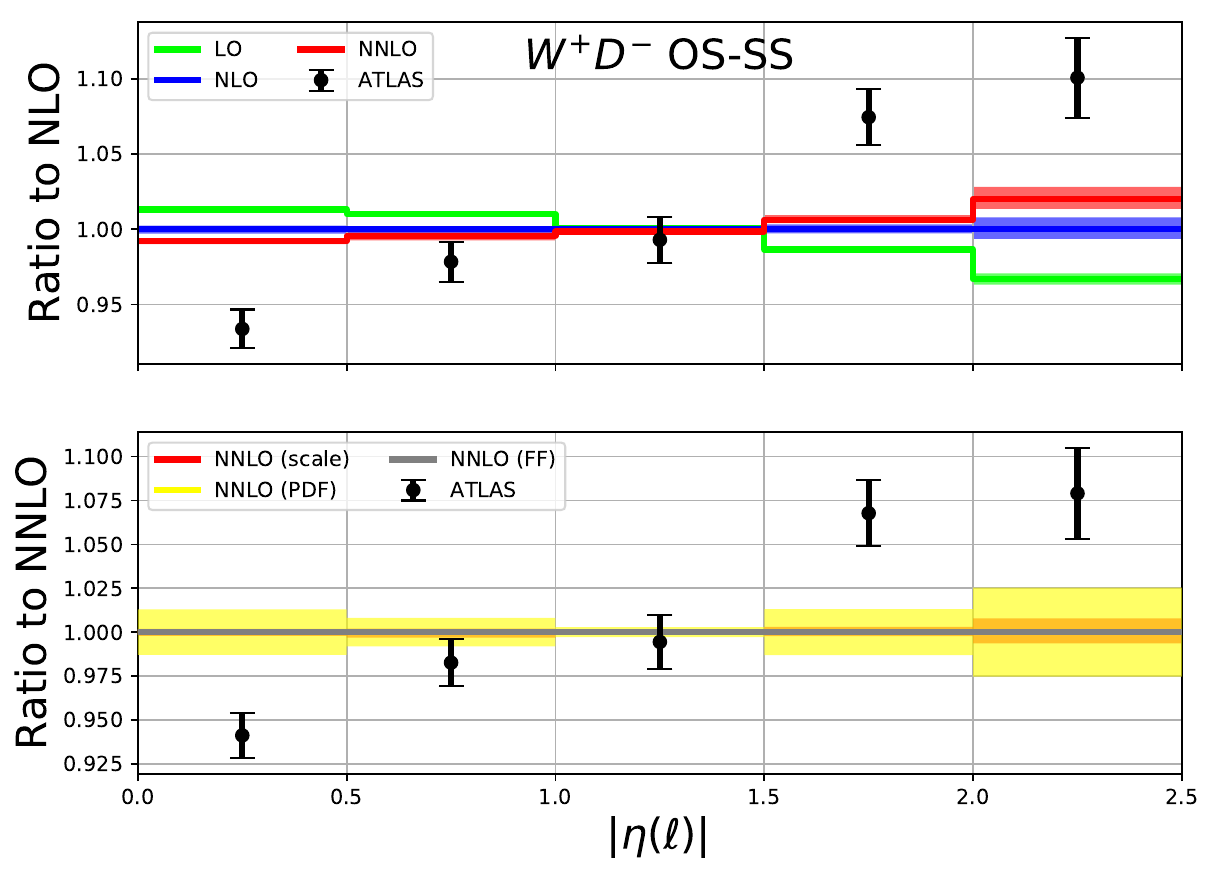}
\includegraphics[width=0.50\textwidth]{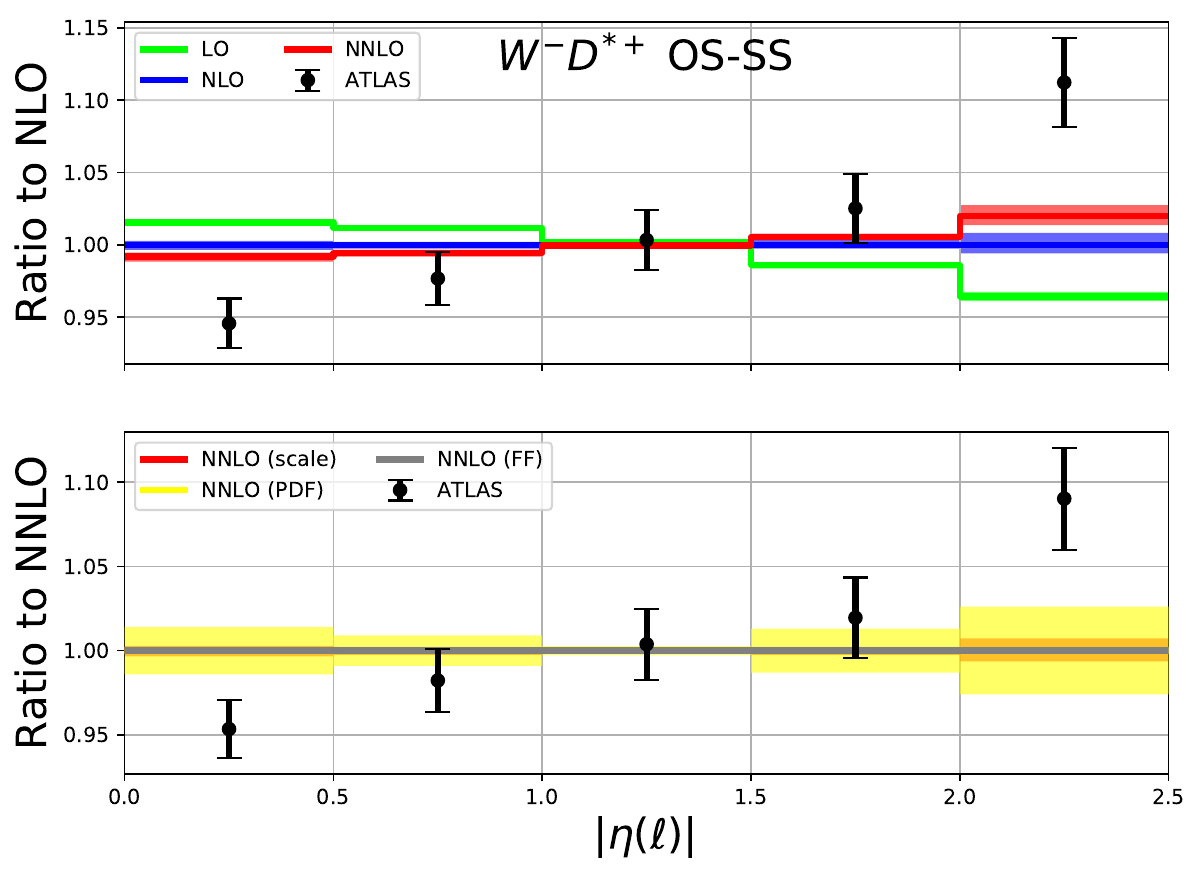}%
\includegraphics[width=0.50\textwidth]{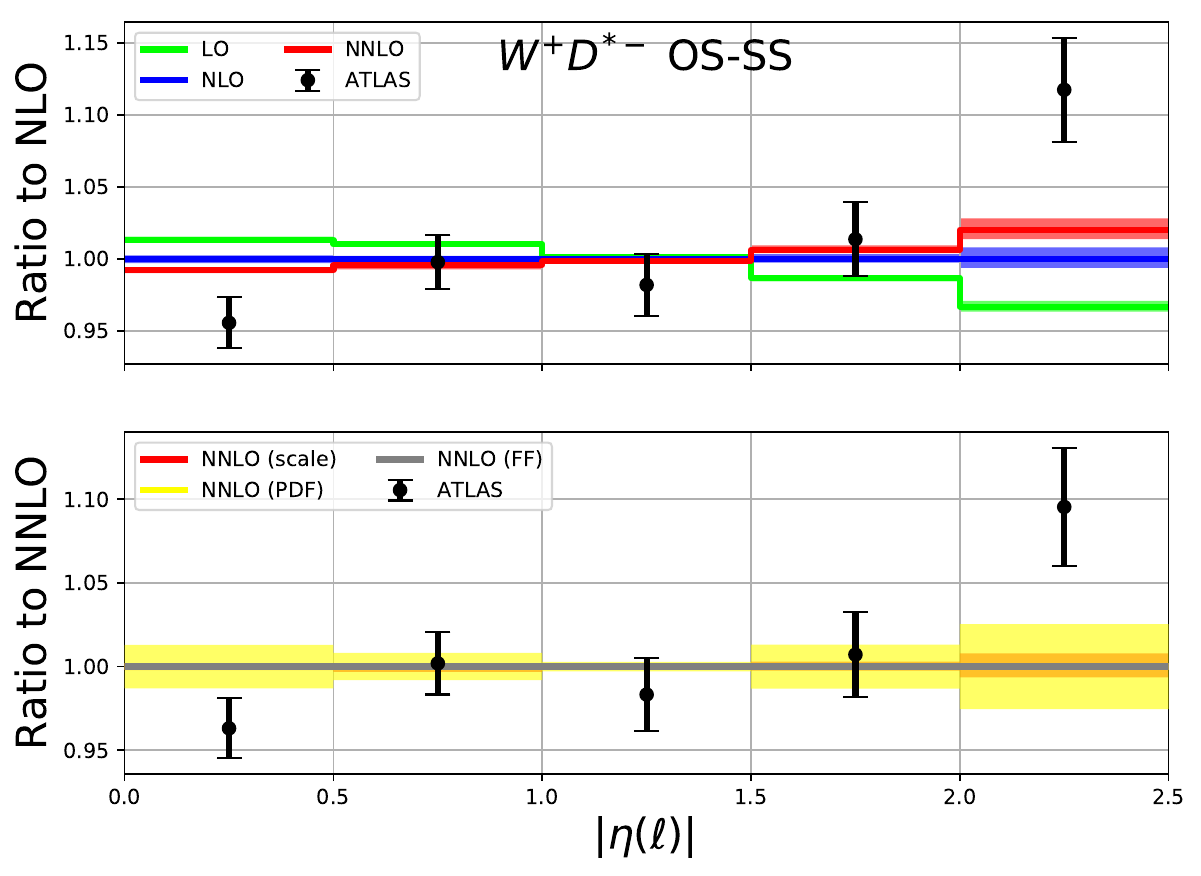}
\caption{As in Figure \ref{fig:rapidity}, but for the normalized distributions.}
\label{fig:normalized-rapidity}
\end{figure}

The same kinematical distributions, but normalized to the total cross section, are shown in Figure \ref{fig:normalized-rapidity}.
The uncertainties from missing higher orders, as well as the FFs, cancel out in this ratio, leaving uncertainties in the sub-percent range, independent of the perturbative order.
This representation highlights that the shape of this observable is relatively independent of the scale and that the perturbative corrections to it are minor, albeit improving the data-to-theory ratio slightly.
Further, it emphasizes the aforementioned impact of the PDFs on the shape of the distribution, which indeed could compensate for the observed shape differences within a 95\% confidence interval.
Further, the small sensitivity to the FF motivates the use of the (normalized) observable as an additional input for PDF fits without considering a simultaneous FF fit.

\begin{figure}[t]
\includegraphics[width=0.50\textwidth]{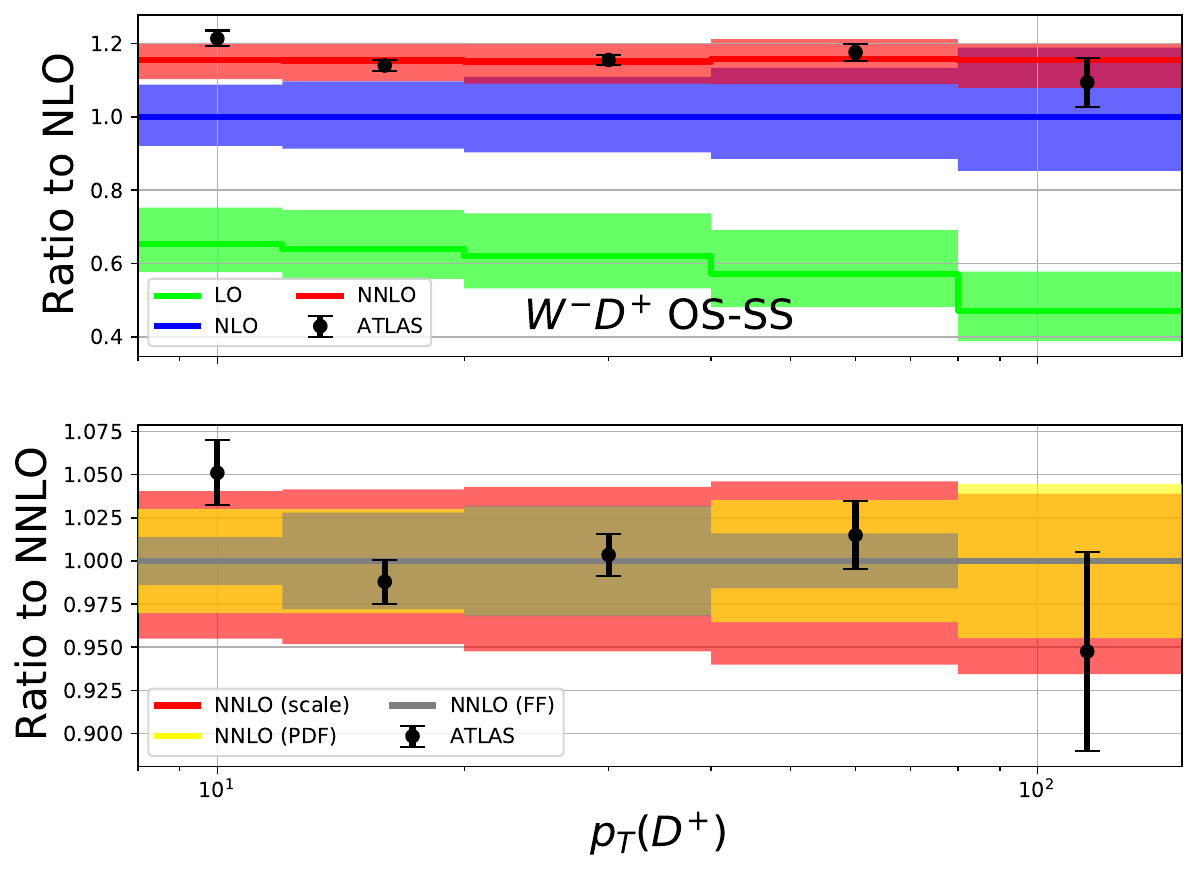}%
\includegraphics[width=0.50\textwidth]{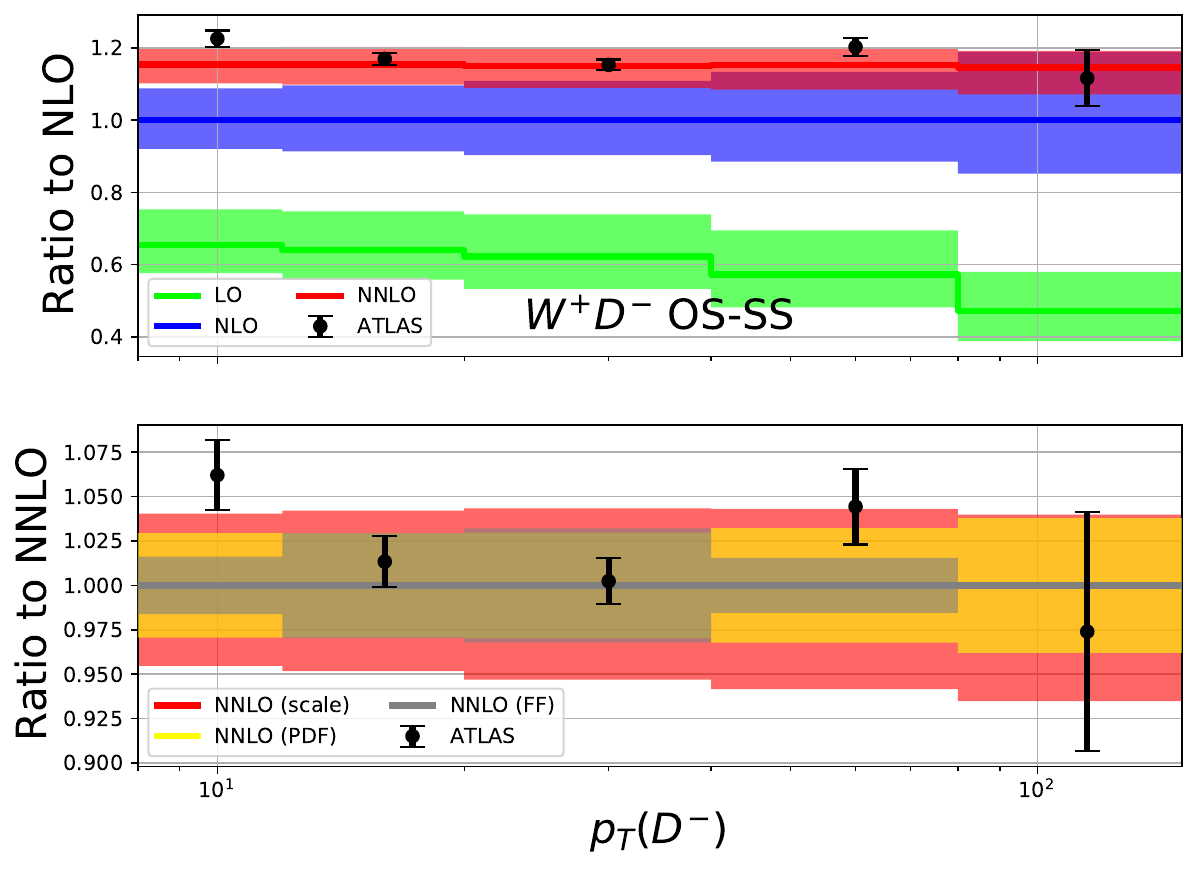}
\includegraphics[width=0.50\textwidth]{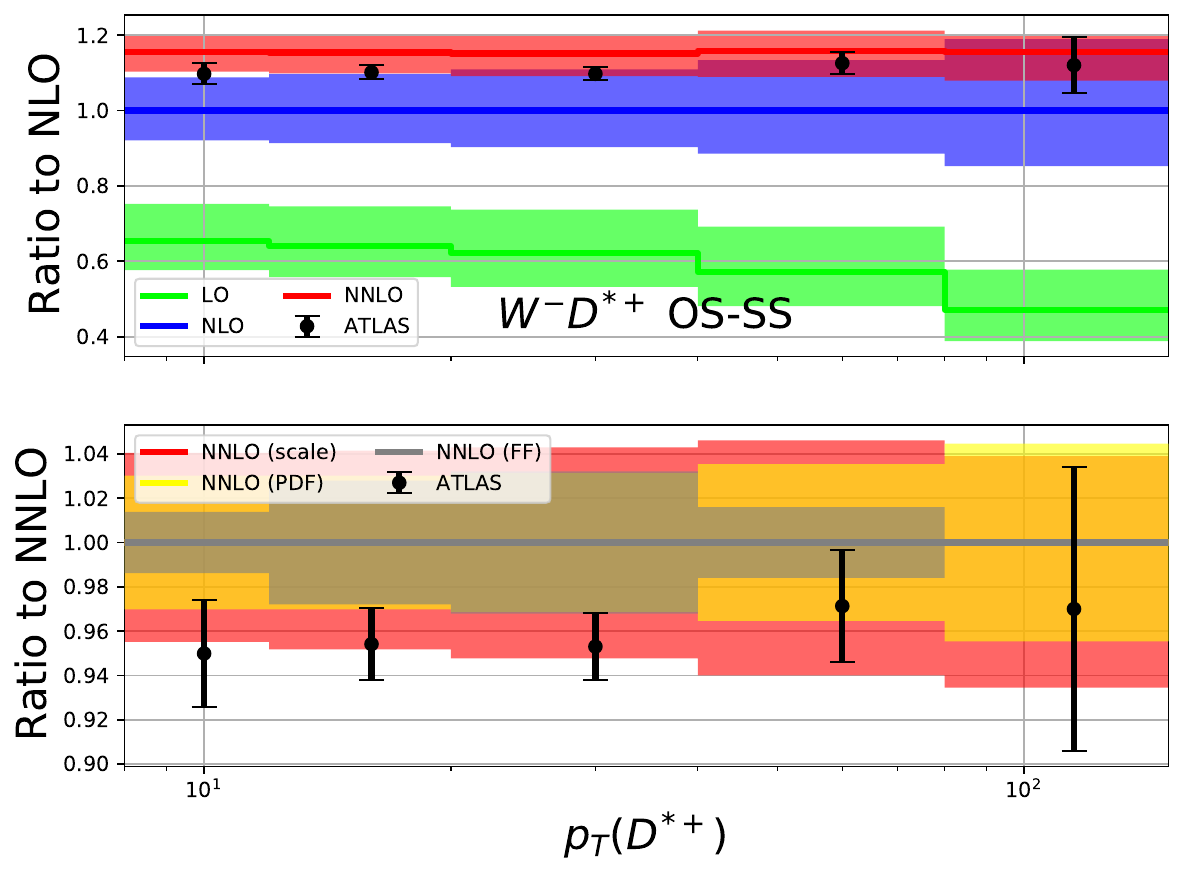}%
\includegraphics[width=0.50\textwidth]{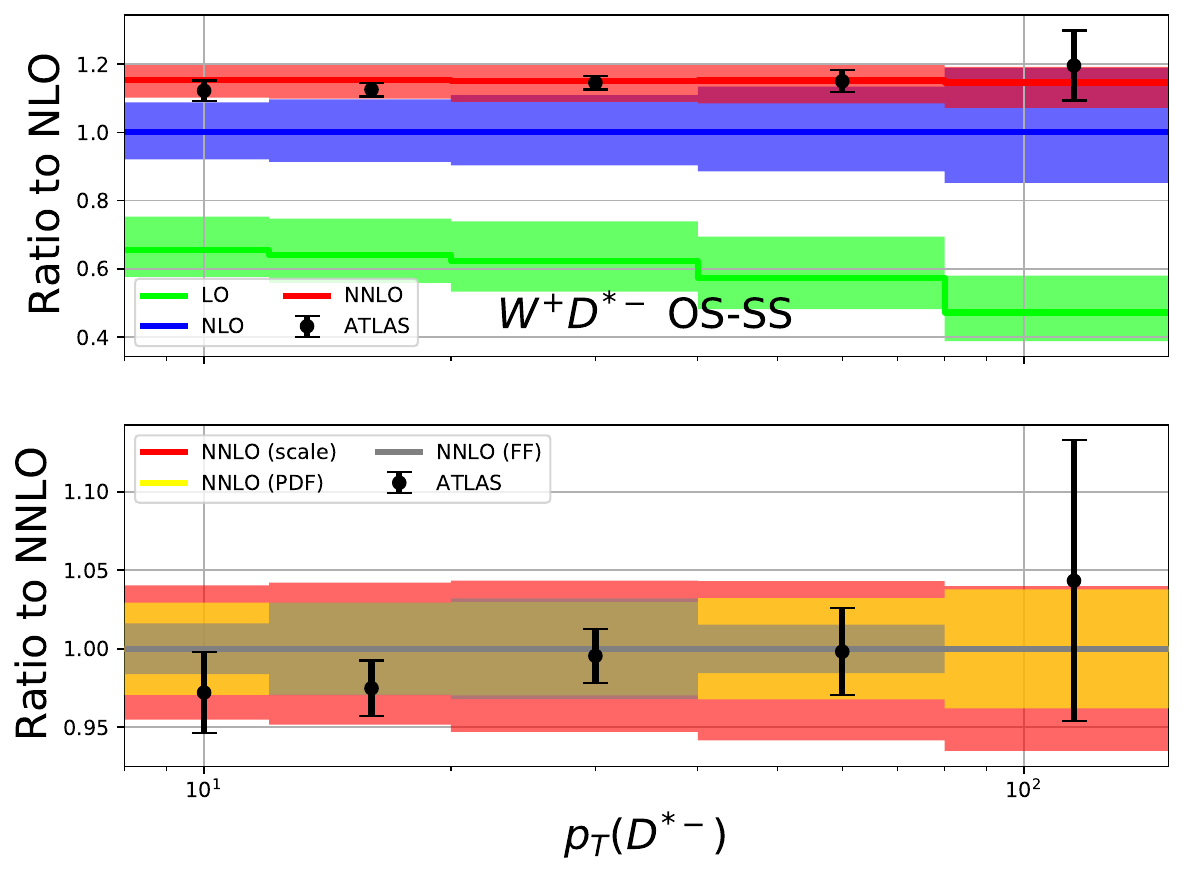}
\caption{As in Figure \ref{fig:rapidity}, but for the $D$-meson transverse momentum distributions.}
\label{fig:transverse-momentum}
\end{figure}

The transverse momentum distribution of the $D$-mesons is presented in Figure \ref{fig:transverse-momentum} in the same layout as before.
Similar to the rapidity case, we find large NLO QCD corrections of about $80\%$ with respect to LO QCD.
The NNLO QCD corrections turn out to be around ${\sim} 20\%$ and to reduce the scale dependence to about $\pm 5\%$.
Especially at high transverse momentum, this reduction is substantial with respect to NLO QCD.
The theory description of the data is significantly improved at NNLO QCD, even predicting the shape of the distributions well.
The only exception is the $D^{*+}$ distribution, where a visible overshoot in normalization is observed, which is consistent with the observation made in the rapidity distribution.
In contrast to the rapidity spectrum, the FF impacts the shape of the predictions as one expects, since this observable is differential in the $D$-meson kinematics.

\begin{figure}[t]
\includegraphics[width=0.50\textwidth]{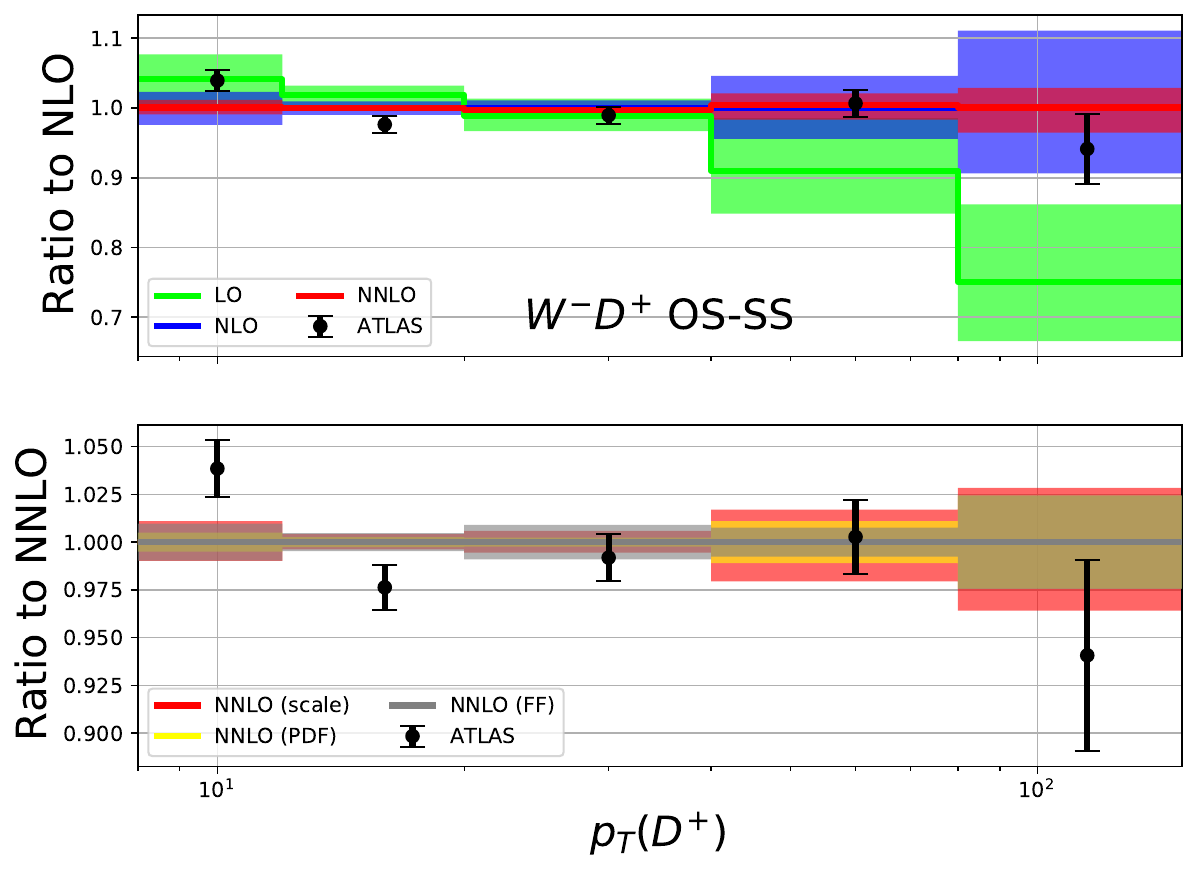}%
\includegraphics[width=0.50\textwidth]{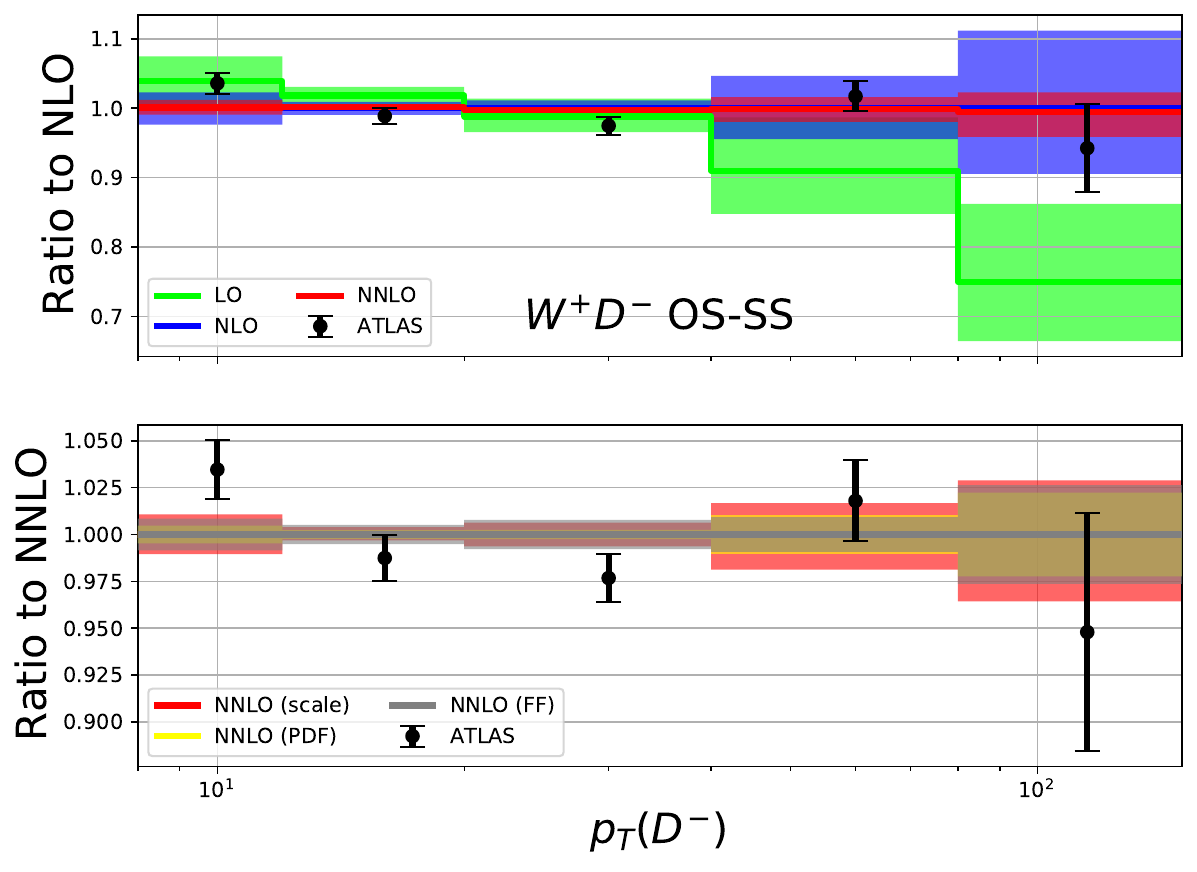}
\includegraphics[width=0.50\textwidth]{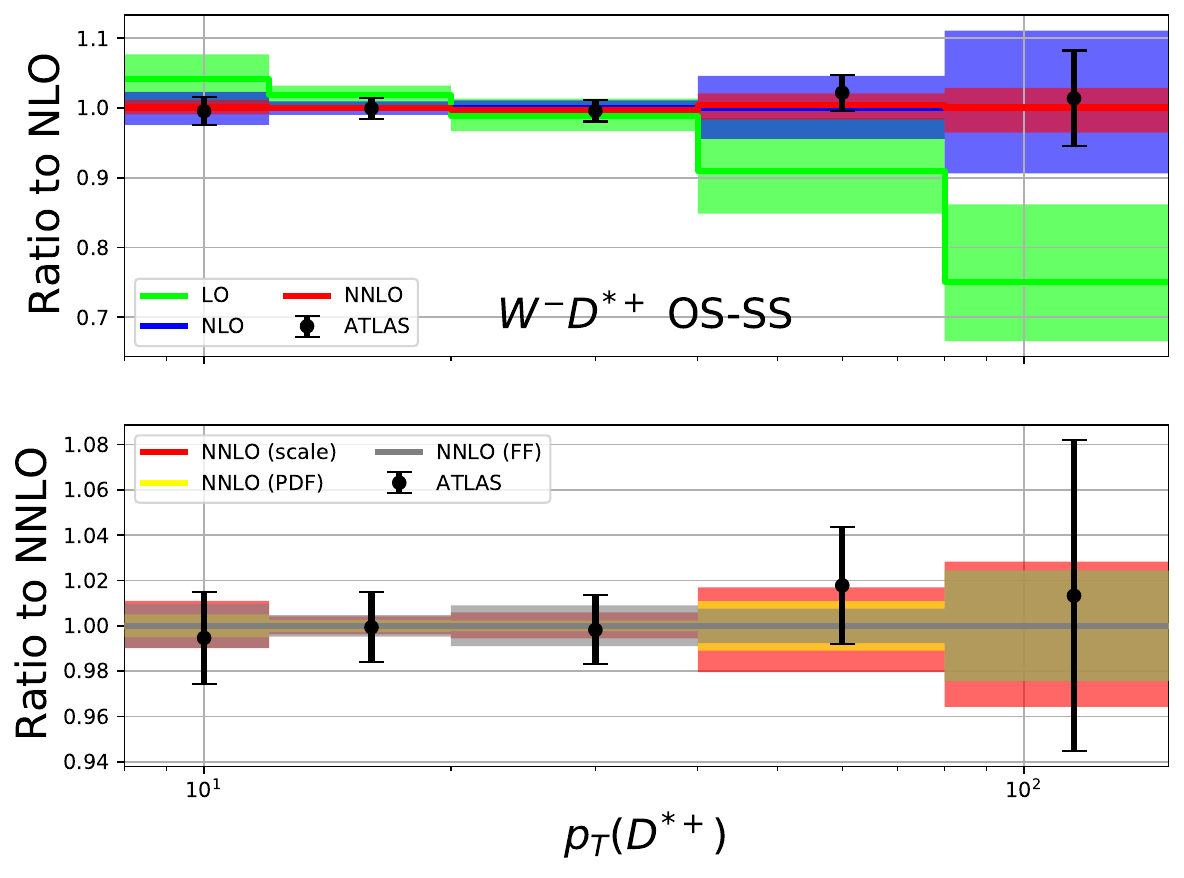}%
\includegraphics[width=0.50\textwidth]{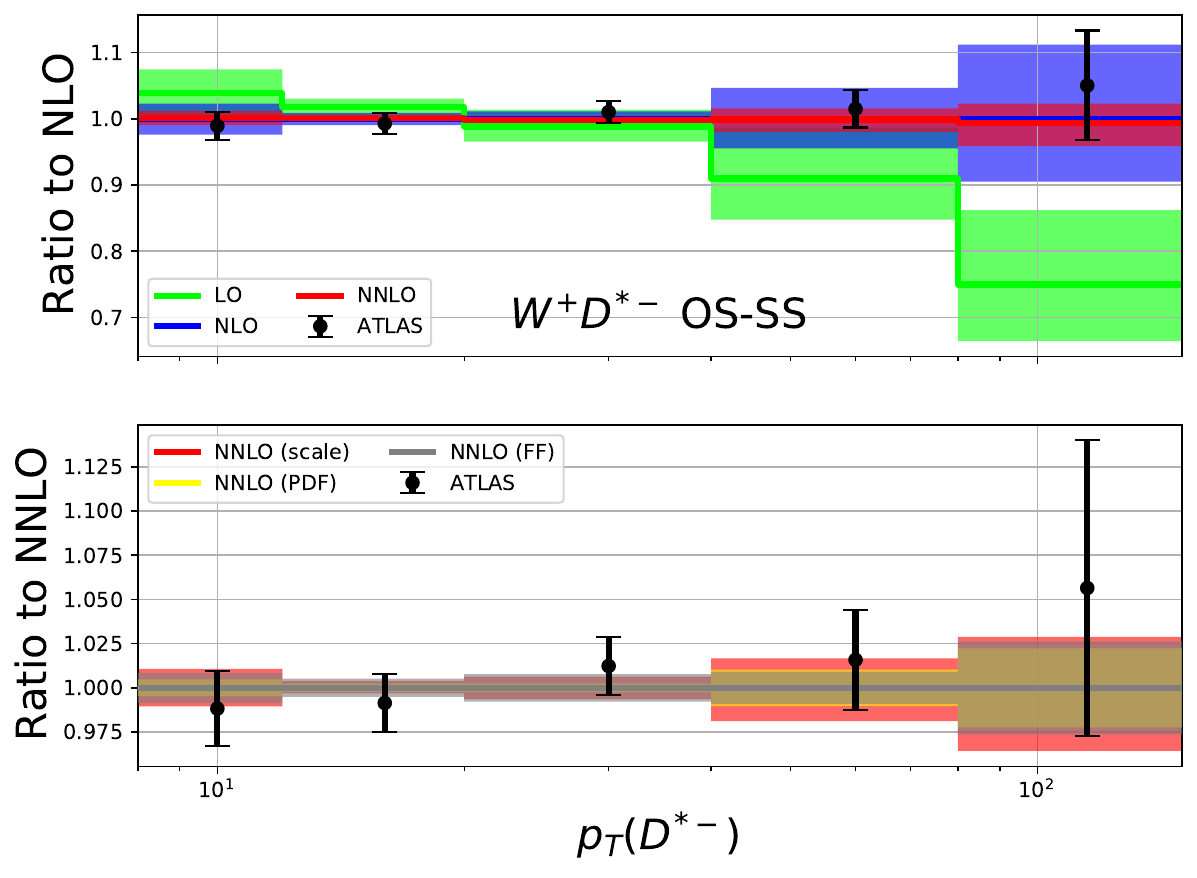}
\caption{As in Figure \ref{fig:transverse-momentum}, but for the normalized distributions.}
\label{fig:normalized-transverse-momentum}
\end{figure}

A striking feature of the normalized transverse momentum spectra in Figure \ref{fig:normalized-transverse-momentum}, besides the well-predicted shapes at NLO and NNLO QCD, is the remaining dependence on the FFs.
The FF uncertainty is of a similar size, if not larger, compared to the scale and PDF dependence.
This observation suggests that a fit of the $D$-meson transverse momentum spectrum and the charged lepton rapidity spectrum would allow a simultaneous extraction of the PDF and FF.

\begin{figure}[t]
\centering
\includegraphics[width=0.49\textwidth]{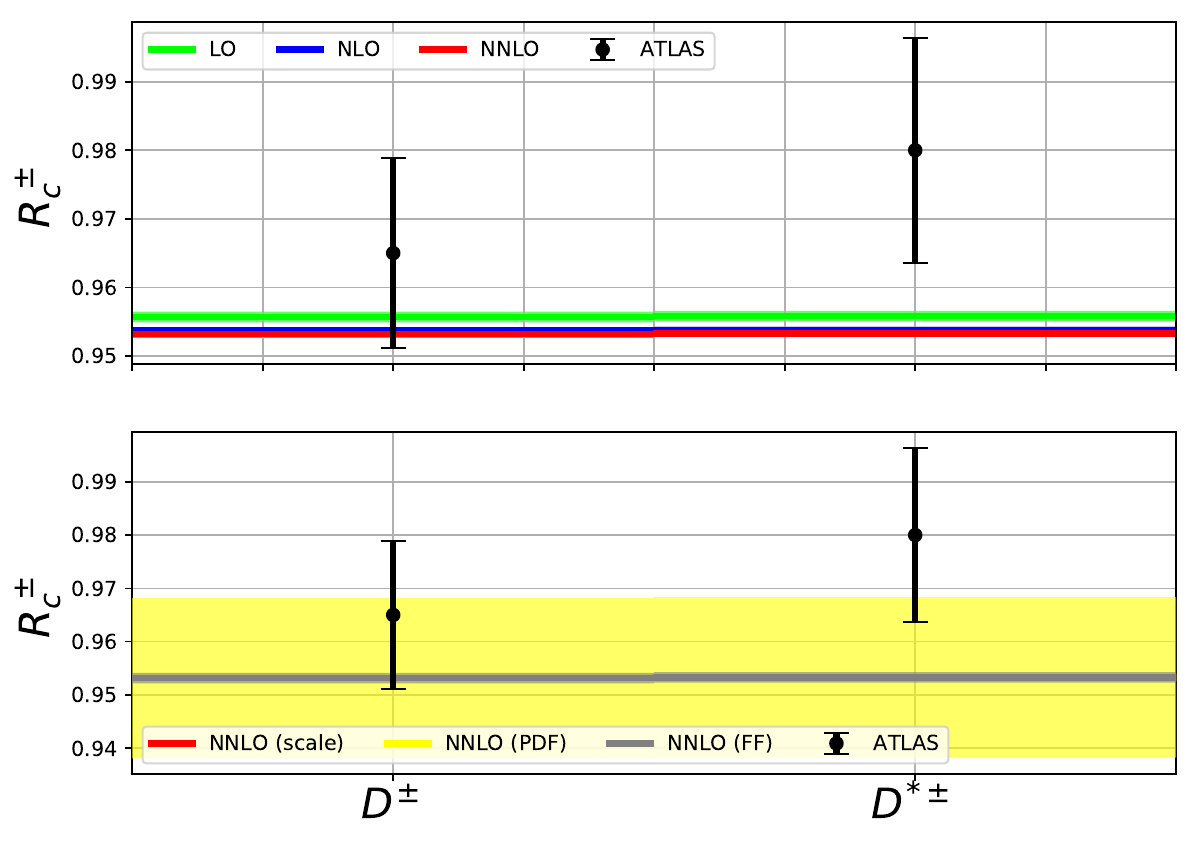}
\caption{As in Figure \ref{fig:rapidity}, but showing the ratio $R^{\pm}_c$ for $D^{\pm}$ and $D^{*\pm}$.}
\label{fig:R}
\end{figure}

The final observable we would like to discuss is the inclusive charge ratios
\begin{align}
    R_c^{\pm}(D) = \frac{\sigma^{\text{OS}-\text{SS}}_\text{fid}(pp \to W^+D)}{\sigma^{\text{OS}-\text{SS}}_\text{fid}(pp \to W^-D)}\quad \text{and}\quad R_c^{\pm}(D^*) = \frac{\sigma^{\text{OS}-\text{SS}}_\text{fid}(pp \to W^+D^*)}{\sigma^{\text{OS}-\text{SS}}_\text{fid}(pp \to W^-D^*)}\;.
\end{align}
The predictions through NNLO QCD are shown in Figure \ref{fig:R} together with ATLAS data.
The higher-order corrections are strongly correlated and thus cancel out in the ratio together with the scale dependence, which we correlated between the numerator and denominator, leading to minimal missing higher-order uncertainties.
The PDF dependence, however, does not cancel, and a substantial uncertainty of about $2\%$ remains, similar to the experimental uncertainties in this case.
Considering the PDF uncertainty, the measurements agree with the NLO and NNLO QCD predictions.

\FloatBarrier
\section{PDF profiling}\label{sec:profiling}

The impact of including the $W^{\pm}D^{(*)\mp}$ data in PDF fits using NNLO calculations can be estimated by PDF profiling. 
Hessian eigenvector sets are evaluated for the $W^{\pm}D^{(*)\mp}$ fiducial cross sections, with each eigenvector set associated with a nuisance parameter $\theta_i$. 
This allows the theory prediction to be expressed as $\vec{T}=\vec{T}(\vec{\theta})$, where $\vec{\theta}$ represents all PDF nuisance parameters of a given set.
Due to the large normalization uncertainties arising from the charm-quark FF, only normalized differential cross sections and cross-section ratios are used in the profiling. 
For each $W^{\pm}D^{(*)\mp}$ channel, five bins of the normalized differential cross section $(1/\sigma)\,d\sigma/d|\eta(\ell)|$, corresponding to the measurement in Ref.~\cite{ATLAS:2023ibp}, are included together with the cross section ratios 
$R^{\pm}_c(D^{(*)})$. 
From the full breakdown of systematic uncertainties provided in Ref.~\cite{ATLAS:2023ibp}, the experimental covariance matrix for the 22 bins (four channels $\times$ five bins for $(1/\sigma)\,d\sigma/d|\eta(\ell)|$, plus $R^{\pm}_c(D)$ and $R^{\pm}_c(D^{*})$) is constructed and denoted as $C_{ij}$.
From these ingredients, a $\chi^2$ function is defined as

\begin{equation} \label{eq:chi2}
\chi^2 = \sum_{ij} (D_i - T_i(\vec{\theta}))\,C^{-1}_{ij}\,(D_j - T_j(\vec{\theta})) 
         + \sum_{i} T_i^2\,\theta_i^2 ,
\end{equation}

where $D_i$ are the central values of the $W^{\pm}D^{(*)\mp}$ measurement, $T_i(\vec{\theta})$ are the corresponding theory predictions for a given set of PDF nuisance parameters $\vec{\theta}$, $C_{ij}$ is the experimental covariance matrix, and $T_i$ are the tolerance factors used in PDF fits to inflate $\Delta\chi^2$ and ensure consistent agreement between all input data sets in cases where the overall $p$-value would otherwise be too low.

The profiling impact is assessed for two PDF sets: \textsc{MSHT20nnlo\-\_as118}~\cite{Bailey:2020ooq} and \textsc{ATLASpdf21\_T1}~\cite{ATLAS:2021vod}. The \textsc{MSHT20nnlo\-\_as118} PDF set was determined with a global PDF fit and has 32 eigenvector sets. Furthermore, it uses dynamic tolerance factors $T_i$, where the tolerance depends both on the nuisance parameter index and sign of the nuisance parameter; i.e. $T_i = T_i(\theta_i/|\theta_i|$), giving $T^{+}_i$ for positive variations of $\theta_i$ and $T^{-}_i$ for negative variations. The \textsc{MSHT20nnlo\-\_as118} PDF fit already includes the $\SI{7}{\tera\electronvolt}$ $W{+}c$ data~\cite{CMS:2013wql} to constrain the $s$ and $\bar{s}$ PDF, and together with tolerance factors that need to be used in profiling, the impact of the new $\SI{13}{\tera\electronvolt}$ ATLAS $W^{\pm}D^{(*)\mp}$ data is not expected to be too large.

On the other hand, \textsc{ATLASpdf21\_T1} PDF set was determined from a fit to mostly ATLAS data together with the deep inelastic scattering data from HERA. As such, it does not include any $W{+}c$ data yet. Furthermore, no tolerances are required in profiling the \textsc{ATLASpdf21\_T1} PDF set, therefore, the impact of adding the new $W^{\pm}D^{(*)\mp}$ data is expected to be large. The \textsc{ATLASpdf21\_T1} PDF set has 21 hessian eigenvectors sets, which can be profiled, and 10 additional variations representing the model and parametrization uncertainties. Those 10 variations are not profiled, but are rather included via an additional covariance matrix that is summed up together with the experimental covariance matrix from the $W^{\pm}D^{(*)\mp}$ data.

\begin{figure}[htpb]
\centering
\includegraphics[width=0.50\textwidth]{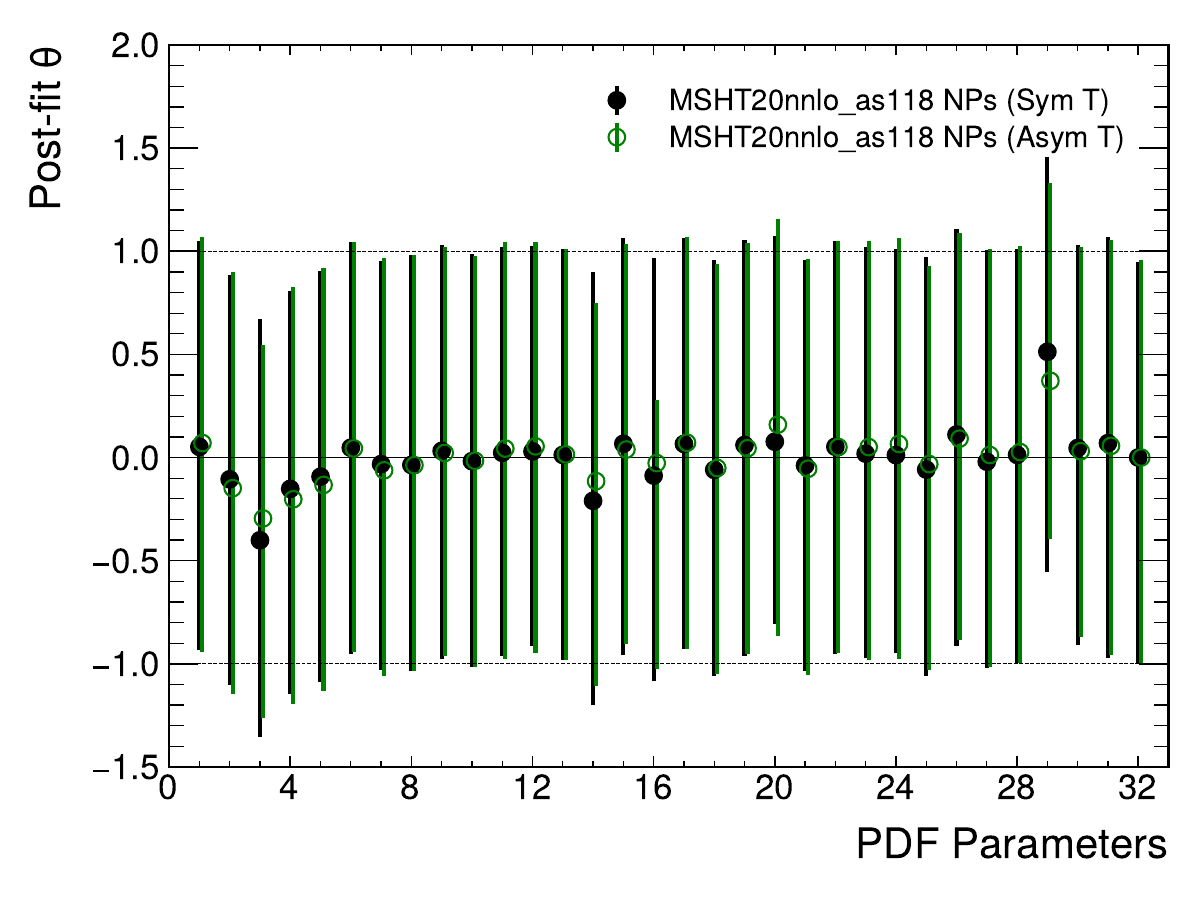}%
\includegraphics[width=0.50\textwidth]{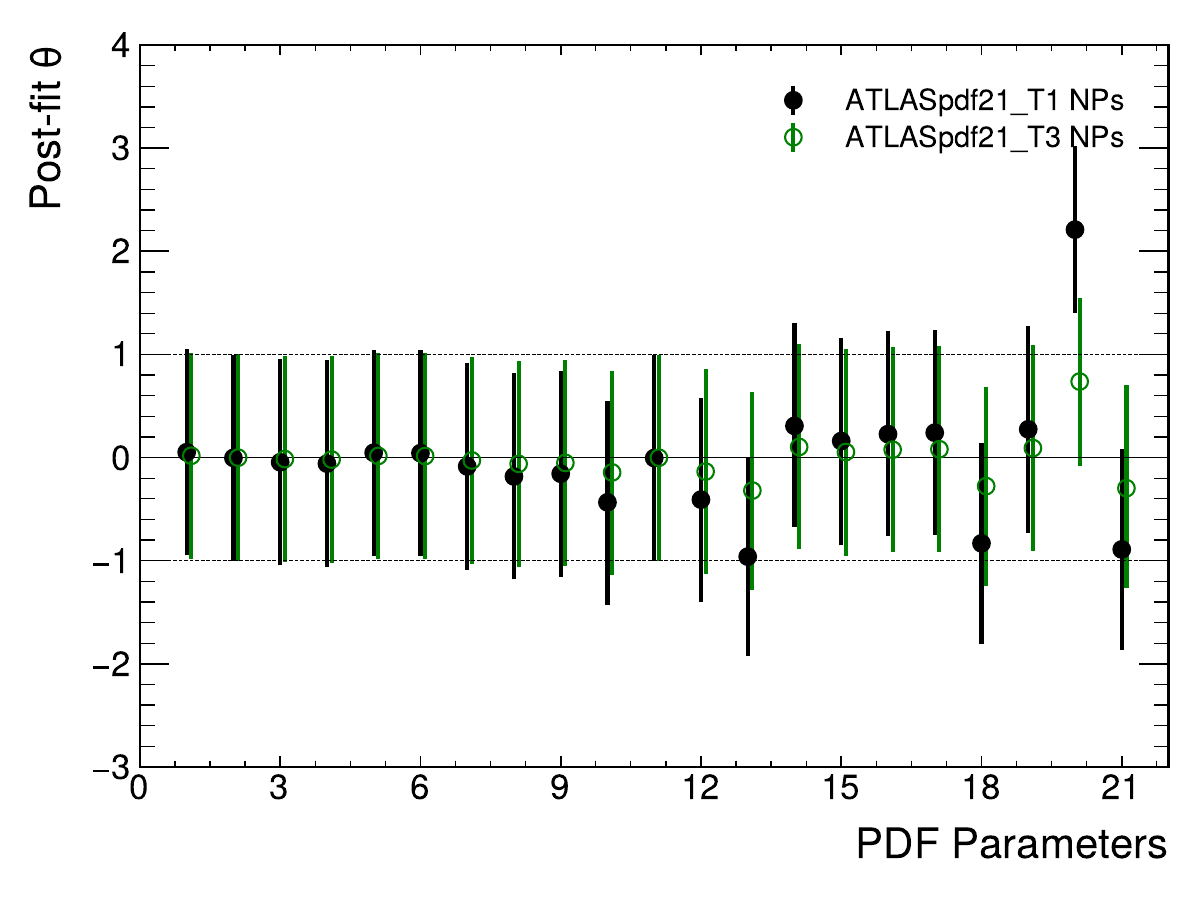}
\caption{Post-fit PDF nuisance parameter pulls and constraints for the \textsc{MSHT20nnlo\-\_as118} (top) and \textsc{ATLASpdf21} (bottom) PDF sets. For the \textsc{MSHT20nnlo\-\_as118} PDF set profiling was performed both for asymmetric tolerances (green points) and symmetrized tolerances (black points).
For \textsc{ATLASpdf21} PDF set both the \textsc{ATLASpdf21\_T1} (black points) and \textsc{ATLASpdf21\_T3} (green points) variants were profiled.
}
\label{fig:profiling:pulls}
\end{figure}

The resulting post-fit PDF nuisance parameter values and their uncertainties are shown in Figure~\ref{fig:profiling:pulls}. For \textsc{MSHT20nnlo\_as118} two profiling strategies were tested: asymmetric tolerance factors with $T^{+}_i$ and $T^{-}_i$ and symmetric tolerances where $T_i$ is defined as $T_i=(T^{+}_i + T^{-}_i)/2$. Similar trends are observed for both strategies and nuisance parameters pulls differ by at most 20\% between the two. For \textsc{MSHT20nnlo\-\_as118}, the largest pulls are observed in nuisance parameters $\theta_3$ and $\theta_{29}$, with their primary parameters being $\alpha_{S,6}$ and $\eta_{s+}$~\cite{Bailey:2020ooq}, corresponding to sea quark PDF and large-$x$ strangeness respectively. These pulls indicate that the additional $W^{\pm}D^{(*)\mp}$ data brings extra sensitivity to the strange quark PDFs, as expected. Several large pulls are observed when profiling the \textsc{ATLASpdf21\_T1} PDF set, ranging up to a two standard deviation shift, indicating the large impact of the additional data. Additionally, the \textsc{ATLASpdf21\_T3} PDF set was profiled for comparison. Pulls are observed in the same parameters as for the \textsc{T1} PDF, but their magnitudes are smaller, as expected. Nevertheless, a significant impact on both PDF sets is observed.

\begin{figure}[tpb]
\centering
\includegraphics[width=0.50\textwidth]{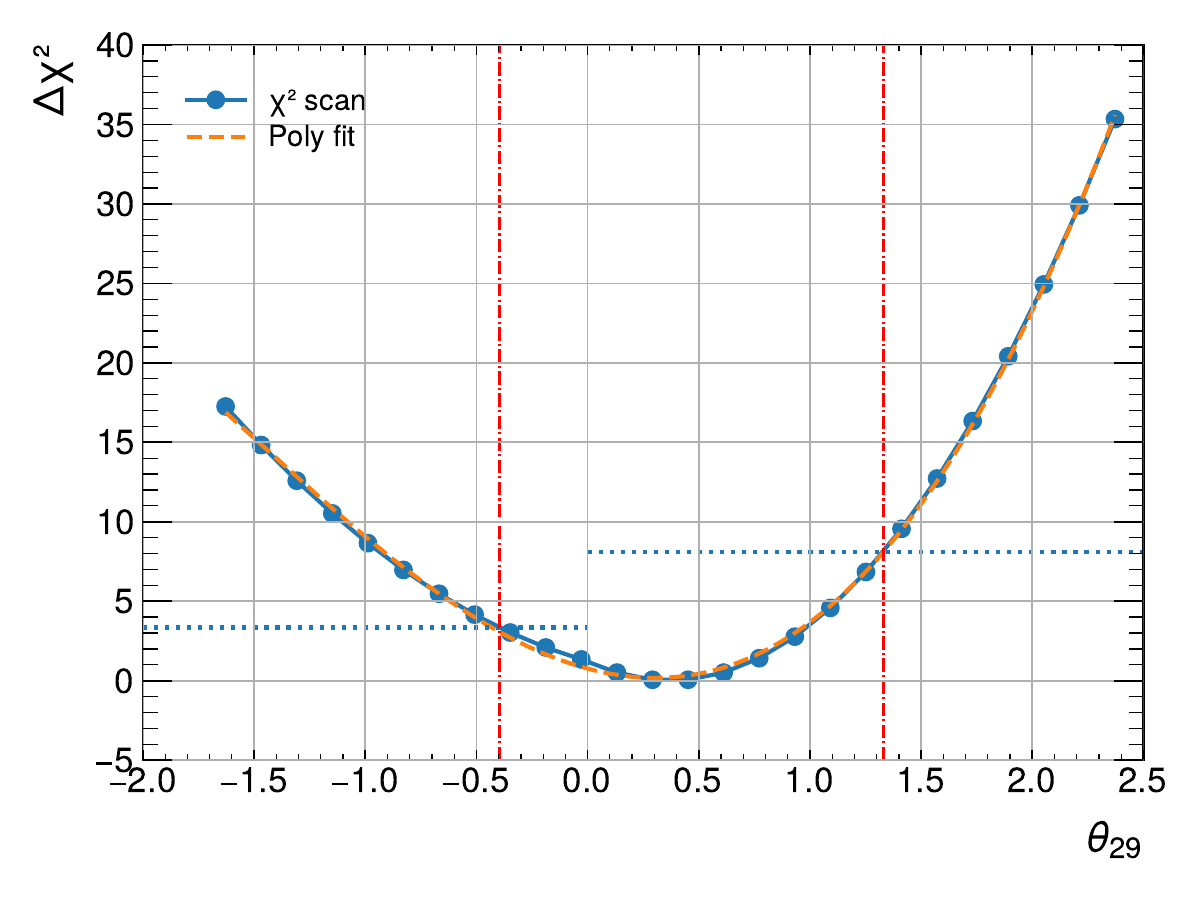}%
\includegraphics[width=0.50\textwidth]{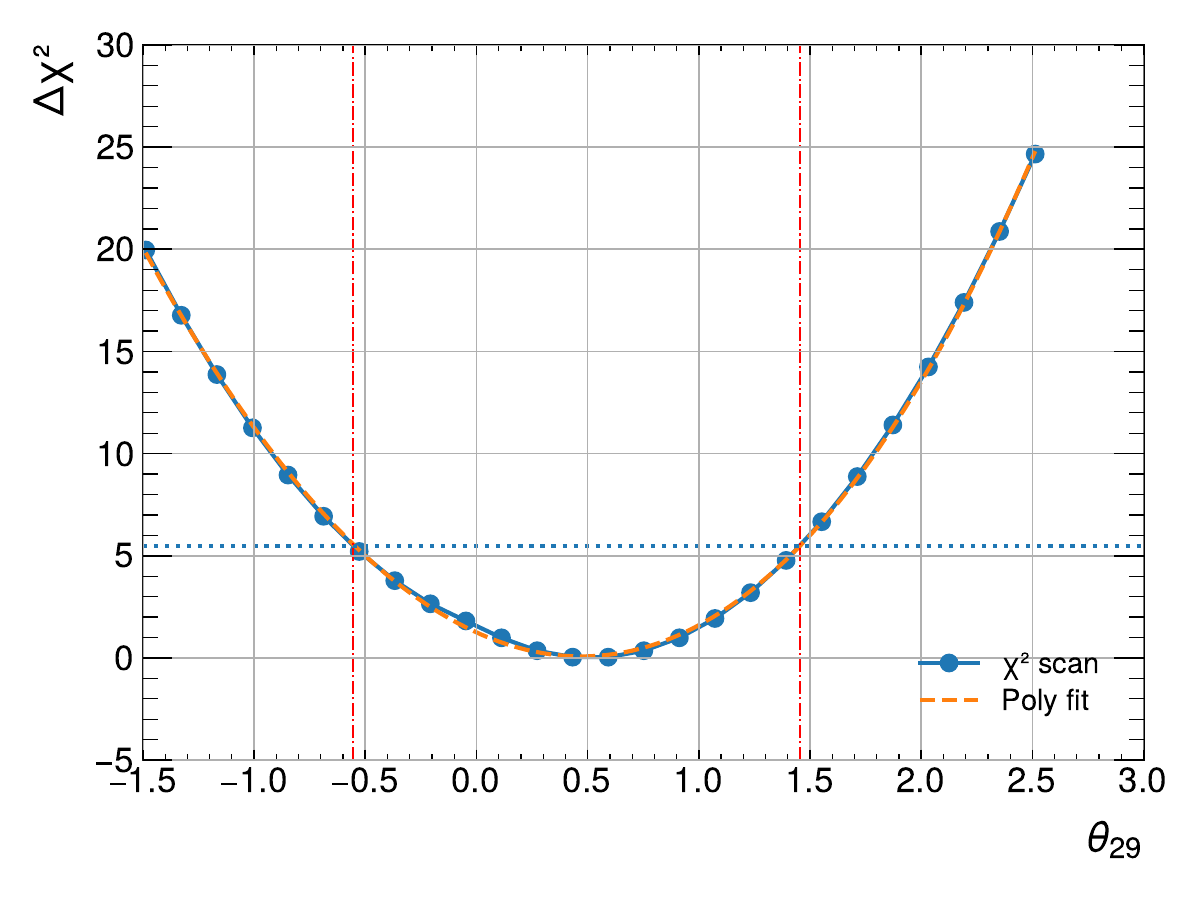}
\caption{Examples of $\Delta\chi^2$ scans for nuisance parameter $\theta_{29}$ of the \textsc{MSHT20nnlo\_as118} PDF set for asymmetric (left) and symmetric (right) tolerance settings. The horizontal dotted blue lines represent the $\Delta\chi^2=T_i^2(\theta_i/|\theta_i|)$ requirement as explained in the text. The vertical dashed red lines represent the one standard deviation uncertainties on the given nuisance parameter after profiling.}
\label{fig:profiling:chi2scan}
\end{figure}

The uncertainties on the post-fit PDF nuisance parameters are determined with a `$\chi^2$ scan', where a scan is performed over each nuisance parameter, fixing it to a certain value, and minimizing the $\chi^2$ formula in Eq.~\eqref{eq:chi2} with respect to the rest of the nuisance parameters. A $\Delta\chi^2$ value is then calculated for each scan point as the difference between the global minimum and the conditional minimum, $\Delta\chi^2=\chi^2(\hat{\vec{{\theta}}})-\chi^2(\hat{\vec{\theta}}_{N-1}|\theta_i=x)$, where $\hat{\vec{{\theta}}}$ indicates the best-fit values of the nuisance parameter from a global minimization and $\hat{\vec{\theta}}_{N-1}$ are the best-fit values for the $N-1$ nuisance parameters given the nuisance parameter $\theta_i$ being fixed to a value $x$. Examples of such $\chi^2$ scans are given in Figure~\ref{fig:profiling:chi2scan} for the nuisance parameter $\theta_{29}$ of the \textsc{MSHT20nnlo\-\_as118} PDF set. The best-fit value of the given nuisance parameter is determined from the minimum of the $\Delta\chi^2$ distribution and its post-fit uncertainty is given by $\Delta\chi^2=T_i^2(\theta_i/|\theta_i|)$. Examples are shown both for the asymmetric and symmetric tolerance cases and the corresponding $T_i^2$ value is given by the dashed blue line. No significant constraints (post-fit nuisance parameter uncertainties less than $\pm1.0$) are observed when profiling the \textsc{MSHT20nnlo\-\_as118} PDF set, while the post-fit uncertainties are reduced down to $\pm0.8$ for the \textsc{ATLASpdf21\_T1} PDF set.

\begin{figure}[tpb]
\centering
\includegraphics[width=0.49\textwidth]{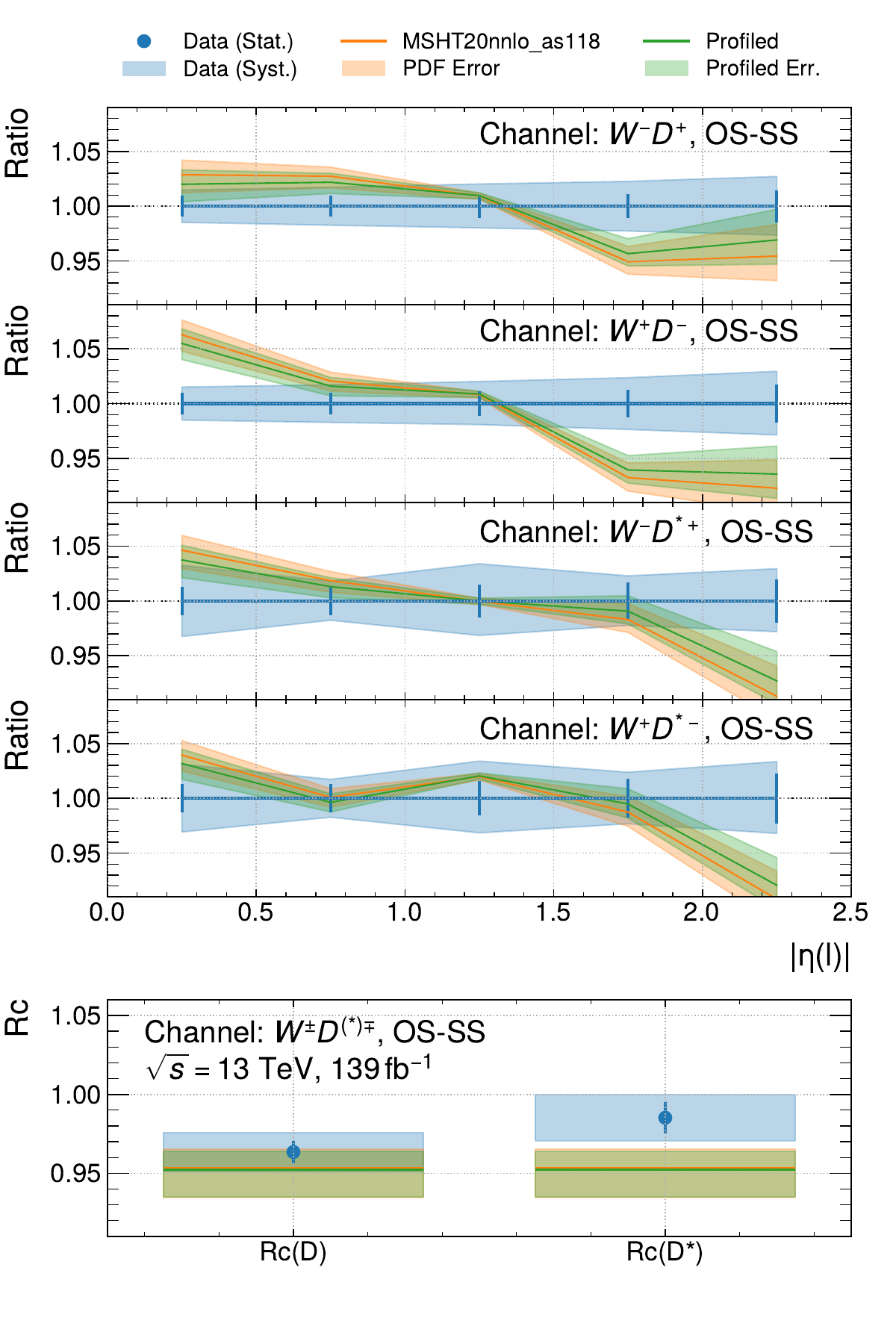}
\includegraphics[width=0.49\textwidth]{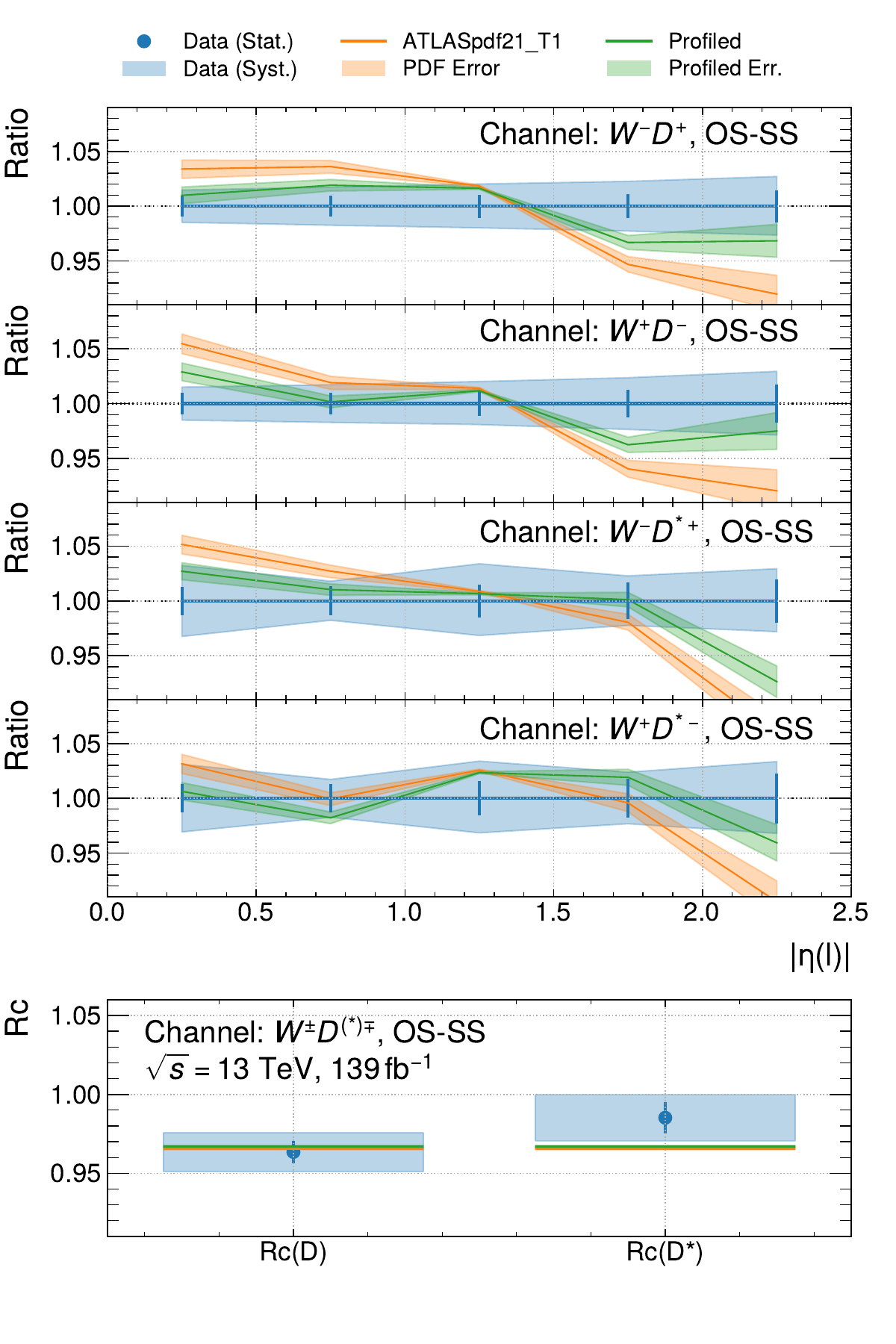}
\caption{Data to prediction ratios of lepton pseudo-rapidity distributions for OS-SS $W^{\pm}D^{(*)\mp}$ channels, and $R^{\pm}_c$ for $D^{\pm}$ (bottom left) and $D^{*\pm}$ (bottom right),  before and after profiling the \textsc{MSHT20nnlo\_as118} (left) and \textsc{ATLASpdf21\_T1} (right) PDF sets.}
\label{fig:profiling:data_mc}
\end{figure}

A comparison of the NNLO theory predictions to the $W^{\pm}D^{(*)\mp}$ data for the baseline and profiled \textsc{MSHT20nnlo\-\_as118} and \textsc{ATLASpdf21\_T1} PDF sets is shown in Figure~\ref{fig:profiling:data_mc}. In all channels, better agreement is observed after profiling. For \textsc{MSHT20nnlo\-\_as118} the total $\chi^2$ before (after) profiling is 31.67 (26.75), corresponding to a $p$-value of $2.4\%$ ($8.4\%$) with 18 degrees of freedom (four channels $\times$ four independent bins for $(1/\sigma)\,d\sigma/d|\eta(\ell)|$, plus $R^{\pm}_c(D)$ and $R^{\pm}_c(D^*)$). For the \textsc{ATLASpdf21\_T1} PDF set the total $\chi^2$ before (after) profiling is 40.37 (21.23), corresponding to a $p$-value of $0.2\%$ ($26.8\%$). Despite the lower initial $p$-value for the \textsc{ATLASpdf21\_T1} PDF set, the post-fit $p$-value is larger than for the \textsc{MSHT20nnlo\-\_as118} PDF set, showing that the PDF can be easier adjusted to the new data.

\begin{figure}[tpb]
\centering
\includegraphics[width=0.49\textwidth]{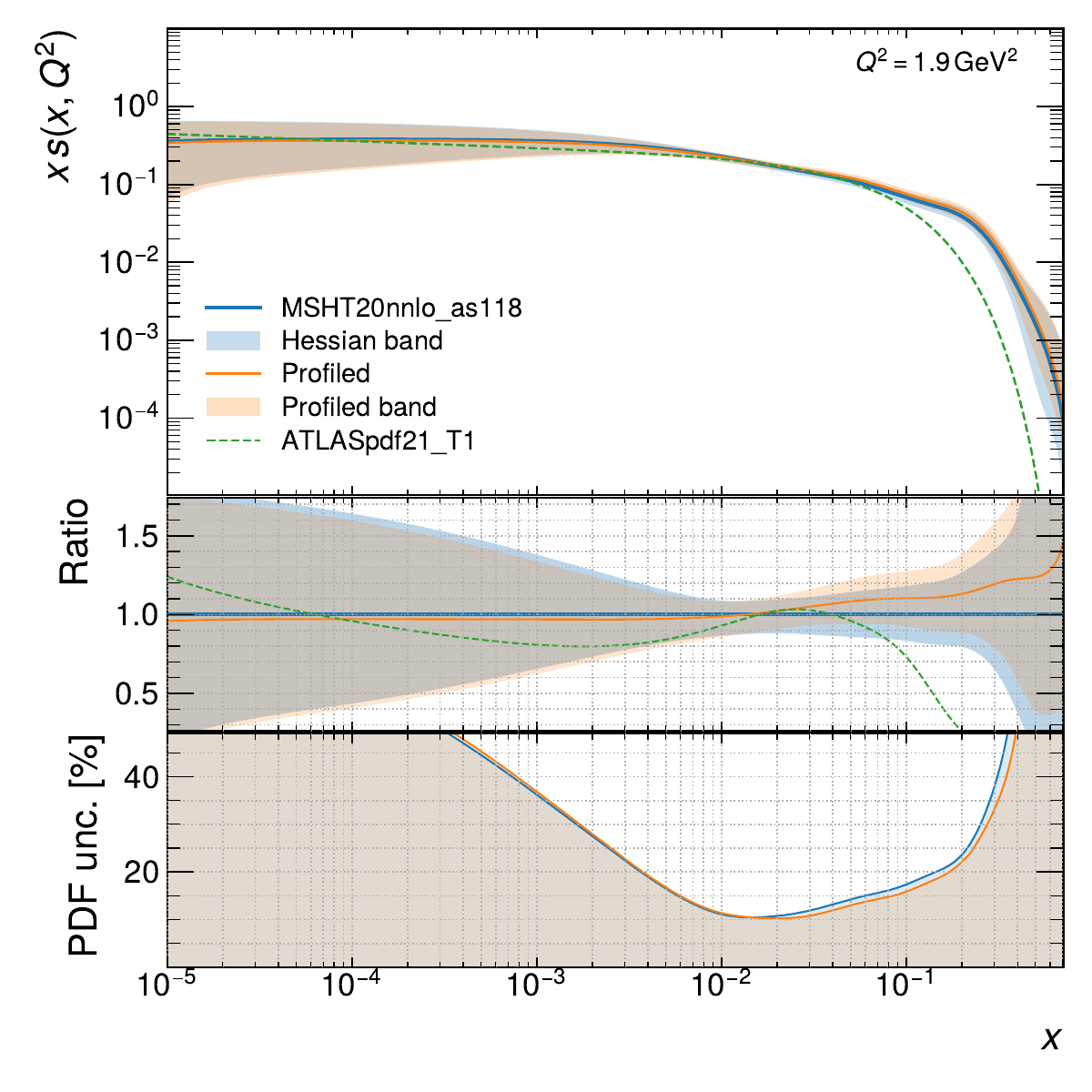}
\includegraphics[width=0.49\textwidth]{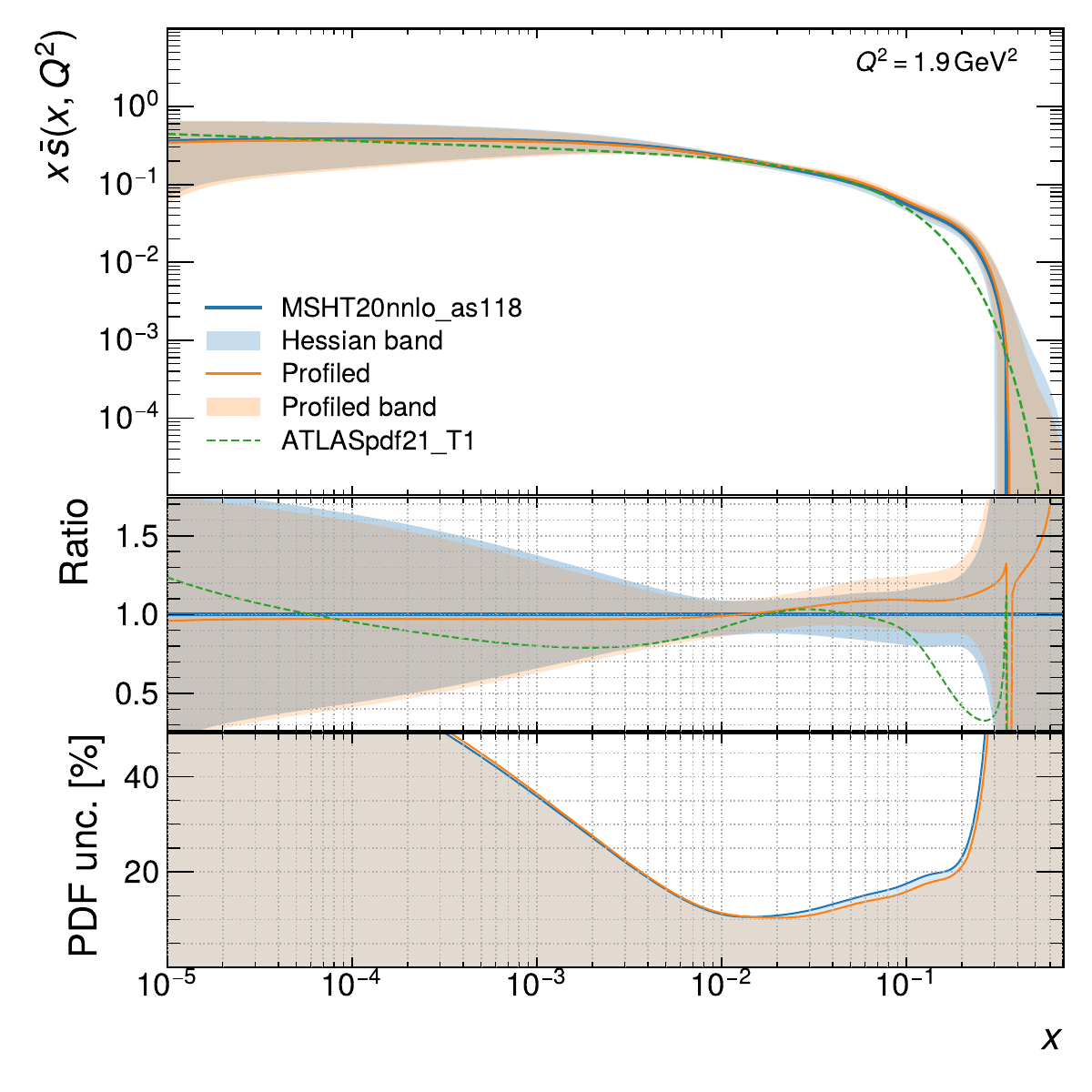}
\includegraphics[width=0.49\textwidth]{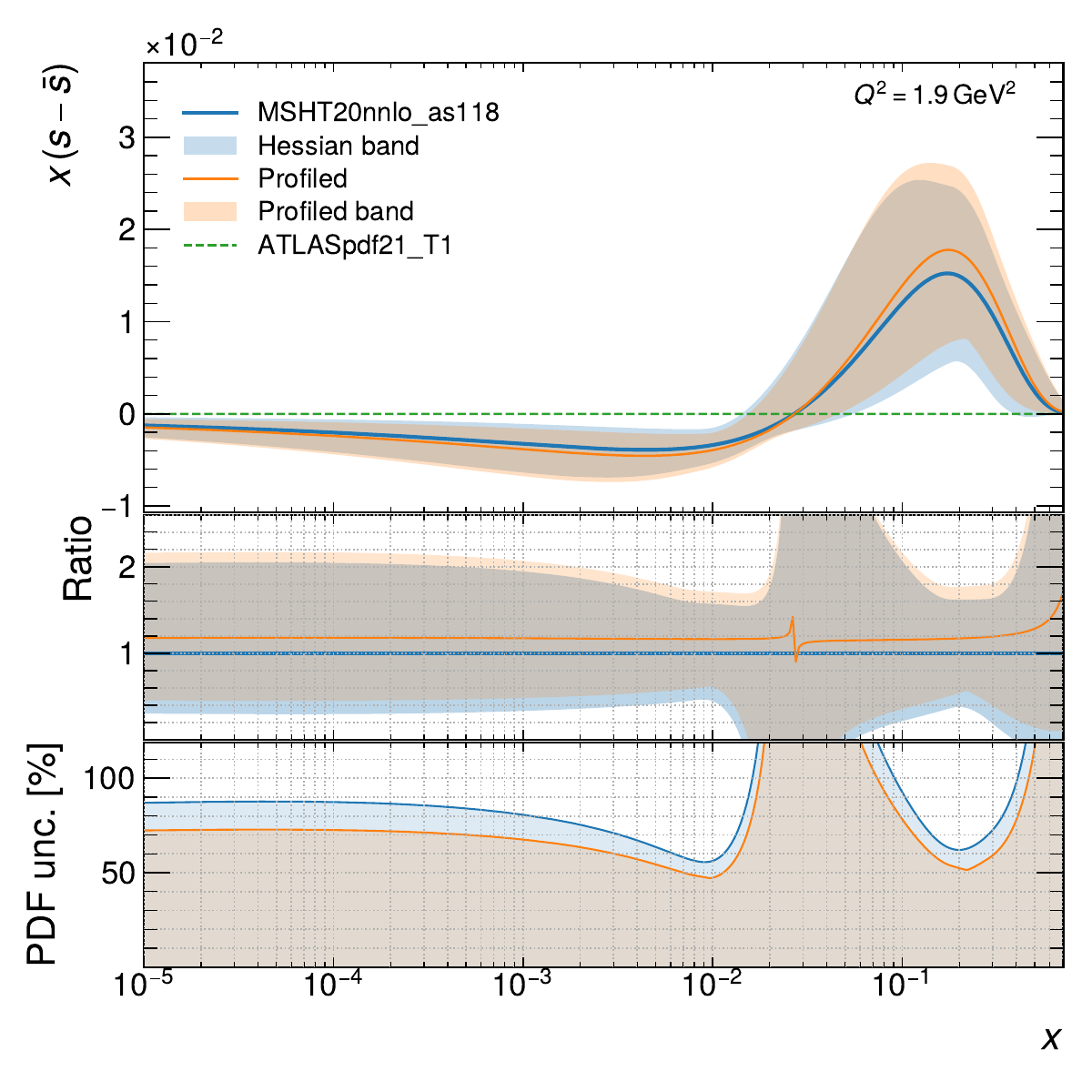}
\includegraphics[width=0.49\textwidth]{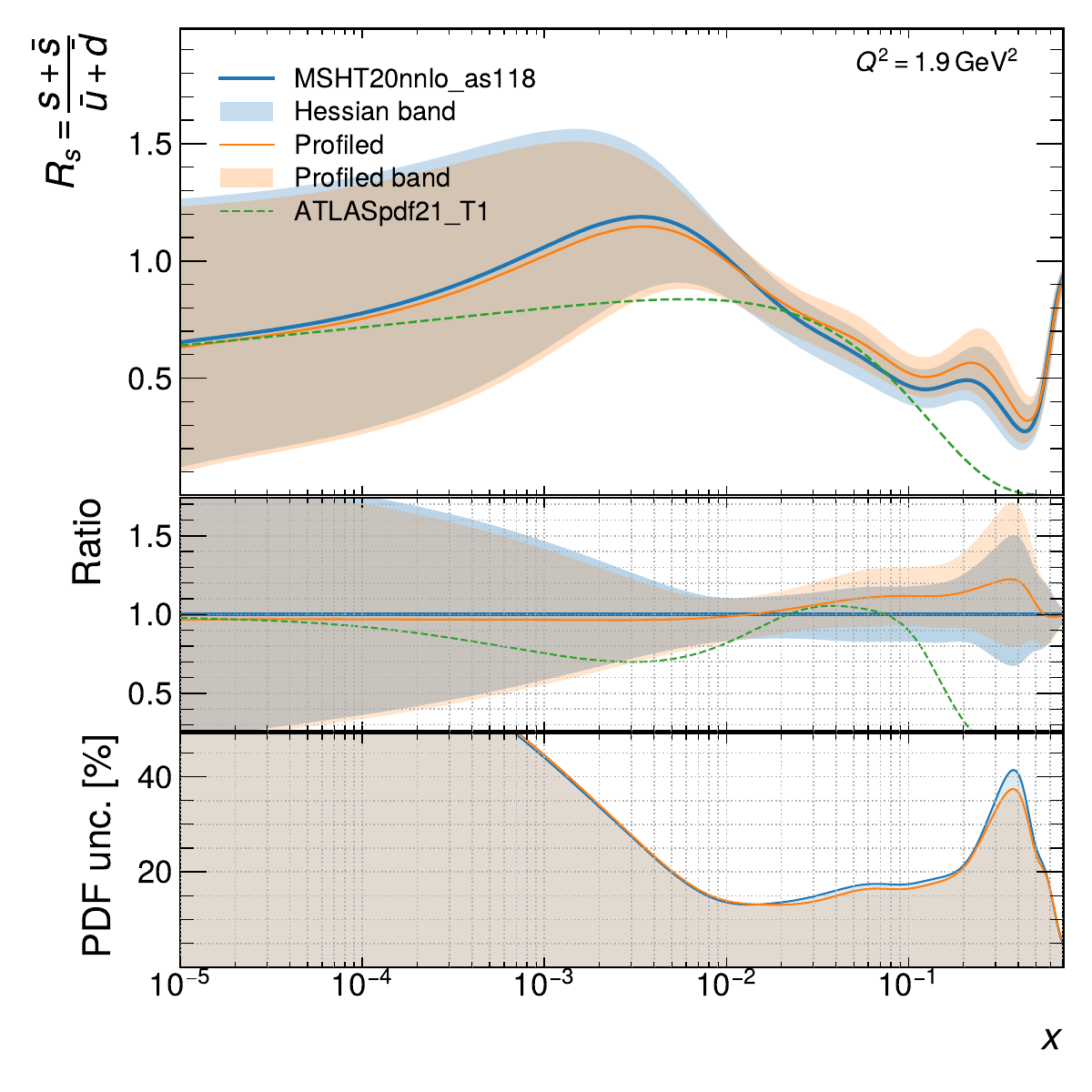}
\caption{The \textsc{MSHT20nnlo\_as118} PDF for $xs(x,Q^2)$ (top left), $x\bar{s}(x,Q^2)$ (top right), strangeness asymmetry $A_s=x(s - \bar{s})$ (bottom left), and strangeness ratio $R_s = (s + \bar{s}) / (\bar{u} + \bar{d})$ (bottom right),  before and after profiling the \textsc{MSHT20nnlo\_as118} PDF set, at $Q^2 = \SI{1.9}{\giga\electronvolt\squared}$. The top panels show the base and profiled PDF distributions with their one standard deviation error bands, the middle panel show the ratio to the base PDF, and the bottom panel shows the relative size of the PDF uncertainty.}
\label{fig:profiling:MSHT20nnlo_as118:PDFs}
\end{figure}
\begin{figure}[htpb]
\centering
\includegraphics[width=0.49\textwidth]{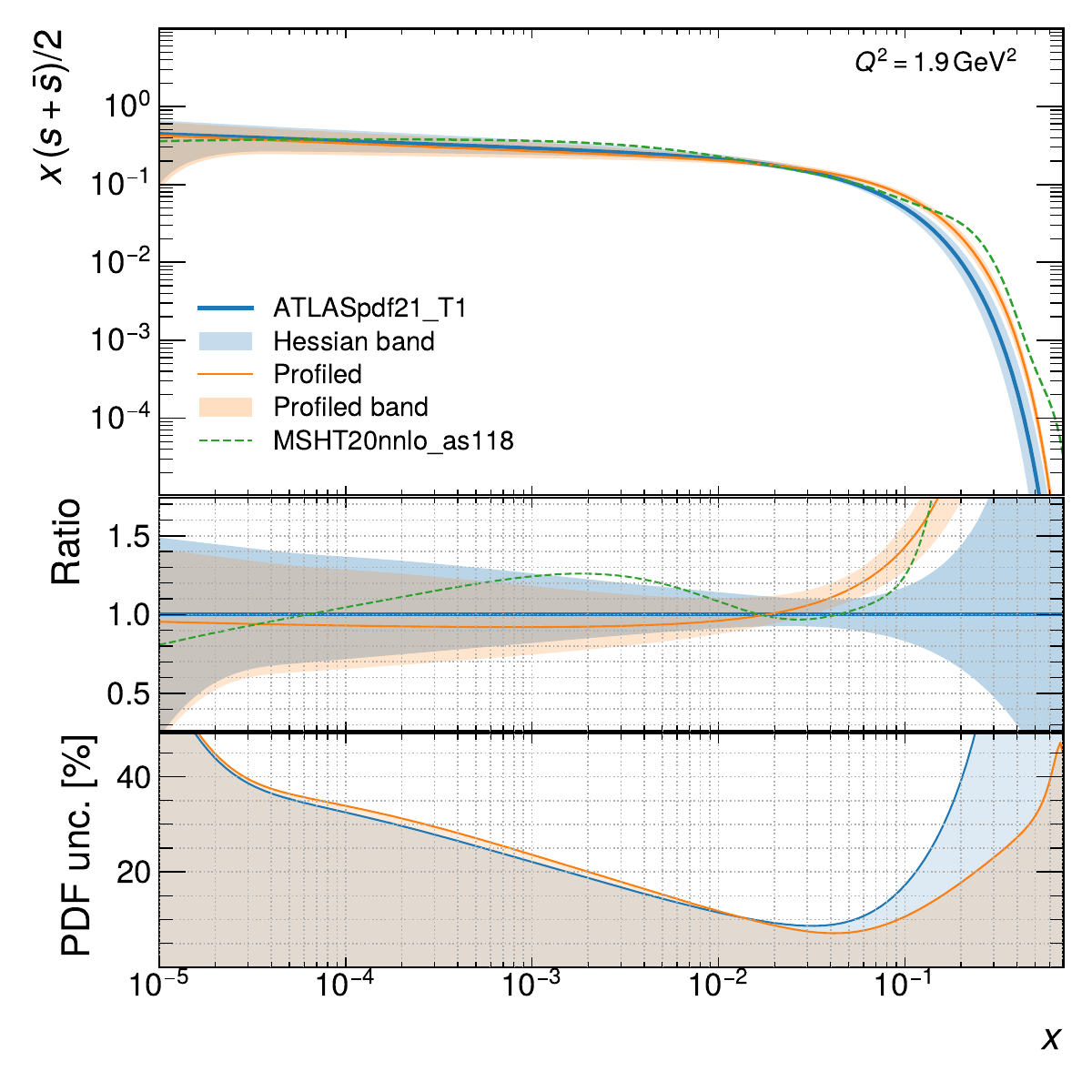}
\includegraphics[width=0.49\textwidth]{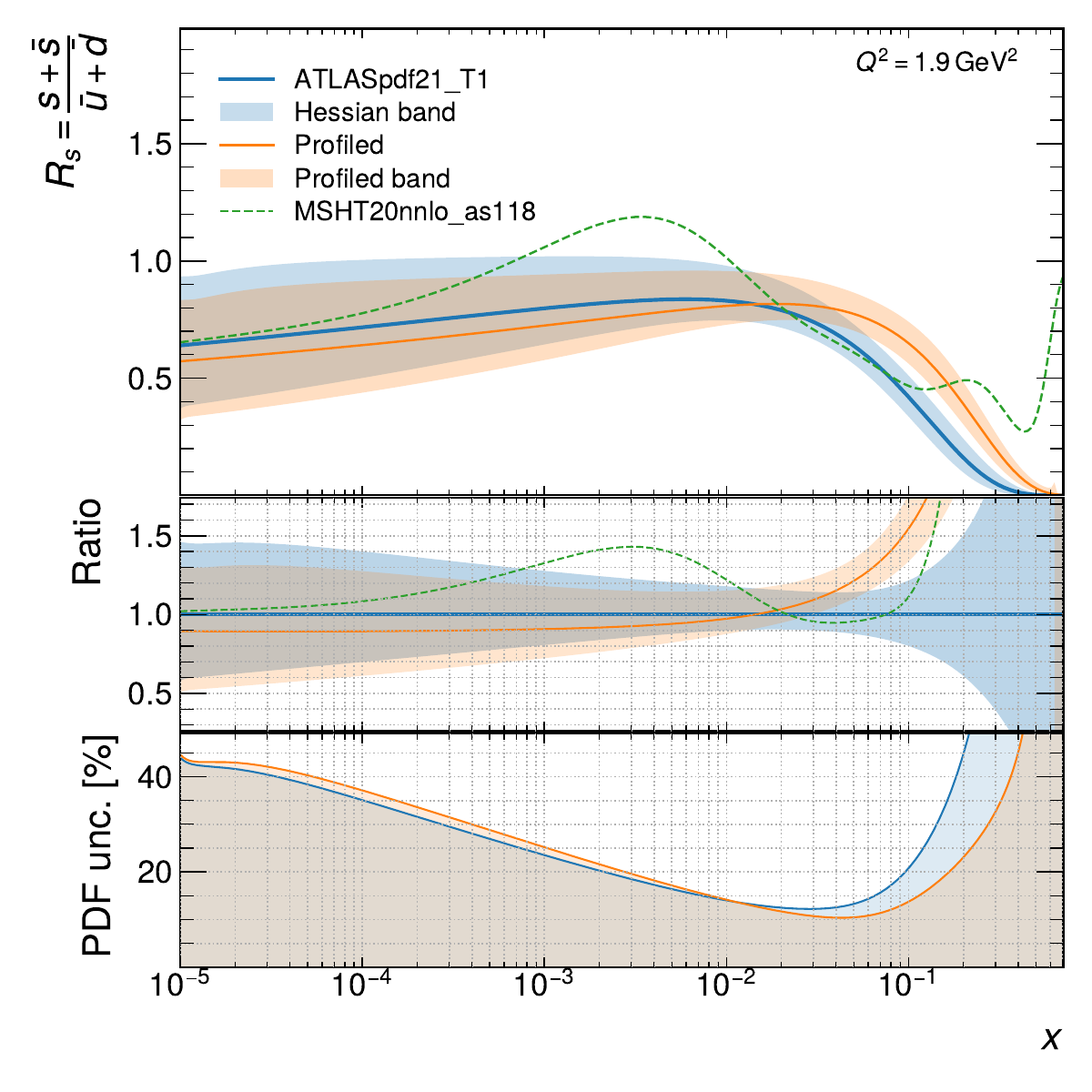}
\caption{The \textsc{ATLASpdf21\_T1} PDF for $x(s + \bar{s}) / 2$ (left) and strangeness ratio $R_s = (s + \bar{s}) / (\bar{u} + \bar{d})$ (right),  before and after profiling the \textsc{ATLASpdf21\_T1} PDF set, at $Q^2 = \SI{1.9}{\giga\electronvolt\squared}$. The top panels show the base and profiled PDF distributions with their one standard deviation error bands, the middle panel show the ratio to the base PDF, and the bottom panel shows the relative size of the PDF uncertainty.}
\label{fig:profiling:ATLASpdf21_T1:PDFs}
\end{figure}

Furthermore, PDF distributions are shown for $xs(x,Q^2)$, $x\bar{s}(x,Q^2)$, strangeness asymmetry $A_s=x(s-\bar{s})$, and the strangeness ratio $R_s=(s+\bar{s})/(\bar{u}+\bar{d})$ for the baseline and profiled \textsc{MSHT20nnlo\_as118} PDF set in Figure~\ref{fig:profiling:MSHT20nnlo_as118:PDFs} and $x(s+\bar{s})/2$ and $R_s$ for the \textsc{ATLASpdf21\_T1} PDF set in Figure~\ref{fig:profiling:ATLASpdf21_T1:PDFs}. Separate $xs(x,Q^2)$ and $x\bar{s}(x,Q^2)$ distributions are not shown for \textsc{ATLASpdf21\_T1} PDF set because it assumes a symmetric $s$ and $\bar{s}$ sea. Shifts of central values up to 20\% are observed for \textsc{MSHT20nnlo\_as118} PDFs, with the largest impact around $x\simeq0.3$ to $x\simeq0.4$. Around the same $x$ values, the relative PDF uncertainties for \textsc{MSHT20nnlo\_as118} are reduced by up to 10\%. On the other hand, large shifts in the \textsc{ATLASpdf21\_T1} PDFs are observed. For comparison the \textsc{MSHT20nnlo\_as118} PDFs are also given in Figure~\ref{fig:profiling:ATLASpdf21_T1:PDFs}, showing that the profiled \textsc{ATLASpdf21\_T1} PDFs move closer to the baseline \textsc{MSHT20nnlo\_as118} PDFs at the medium $x$ range of $x\simeq0.01$ to $x\simeq0.5$.

%

\section{Conclusion}\label{sec:conclusion}

We computed the production of a $W^{\pm}$-boson in association with a $D^{(*)\mp}$-meson at the LHC through NNLO QCD within the collinear factorization approach.
To suppress gluon splitting backgrounds, all observables have been defined as the difference between the cross sections with opposite and the same sign of the meson's electric charge with respect to the $W$-bosons. The phenomenology of NNLO QCD corrections in a typical LHC phase space for two relevant observables, the rapidity of the charged lepton and the transverse momentum of the $D$-meson, has been examined and found to improve the description of recent ATLAS data compared to NLO QCD.

Focusing on the charged lepton rapidity, the NNLO QCD corrections have been found to have a significant impact on the normalization of the cross section, while having a minimal influence on its shape, implying that normalized cross sections are relatively insensitive to these corrections.
Crucially, however, the shape shows a larger dependence on the PDF's functional form, implying constraining power.
Additionally, we studied the impact of non-perturbative fragmentation functions on the rapidity distribution of the charged lepton and found it to be minimal, which in turn justifies the inclusion of this data in a future global PDF fit.
The sensitivity of this observable to the strange-quark PDF has been demonstrated by PDF profiling using recent MSHT and ATLAS PDF sets.
Finally, we studied the transverse momentum distribution of the $D$-mesons, which showed remarkable agreement between NNLO QCD and data.
Not only has the shape of the distributions been improved, but also the remaining estimated theory uncertainties are significantly reduced, specifically at high transverse momentum.
This distribution shows, in contrast to the lepton rapidity, sensitivity to the FFs, which might allow, together with the reduced perturbative uncertainties at NNLO QCD, to constrain them in the future.

Our results confirm that the charged lepton rapidity distribution is a useful input for a PDF fit, further constraining the strange quark content.
They also highlight the possibility of simultaneously extracting the PDF and $D$ fragmentation functions from the combination of two studied observables.
Such a fit would, however, require the computation of FastNLO tables or similar PDF/FF-independent representations of the cross sections, which we leave for future work.

\section*{Acknowledgements}
The authors would like to thank the organizers of the Les Houches PhysTeV25 workshop, on which occasion this project was initiated. T.G.~has been supported by STFC consolidated HEP theory grants ST/T000694/1 and ST/X000664/1. This work was performed using the Cambridge Service for Data Driven Discovery (CSD3), part of which is operated by the University of Cambridge Research Computing on behalf of the STFC DiRAC HPC Facility (www.dirac.ac.uk). The DiRAC component of CSD3 was supported by STFC grants ST/P002307/1, ST/R002452/1 and ST/R00689X/1. M.M.~has been supported by Slovenian Research and Innovation Agency grants J1-50217, N1-0398 and P1-0135.

\bibliographystyle{JHEPmod}
\bibliography{references}

\providecommand{\href}[2]{#2}\begingroup\raggedright\begin{thebibliography}{10}

\bibitem{Lin:2017snn}
H.-W. Lin et~al., {\it {Parton distributions and lattice QCD calculations: a
  community white paper}},  {\em Prog. Part. Nucl. Phys.} {\bf 100} (2018)
  107--160, [\href{http://arxiv.org/abs/1711.07916}{{\tt arXiv:1711.07916}}].

\bibitem{Alekhin:2017kpj}
S.~Alekhin, J.~Bl{\"u}mlein, S.~Moch, and R.~Placakyte, {\it {Parton
  distribution functions, $\alpha_s$, and heavy-quark masses for LHC Run II}},
  {\em Phys. Rev. D} {\bf 96} (2017) 014011,
  [\href{http://arxiv.org/abs/1701.05838}{{\tt arXiv:1701.05838}}].

\bibitem{Hou:2019efy}
T.-J. Hou et~al., {\it {New CTEQ global analysis of quantum chromodynamics with
  high-precision data from the LHC}},  {\em Phys. Rev. D} {\bf 103} (2021)
  014013, [\href{http://arxiv.org/abs/1912.10053}{{\tt arXiv:1912.10053}}].

\bibitem{Bailey:2020ooq}
S.~Bailey, T.~Cridge, L.~A. Harland-Lang, A.~D. Martin, and R.~S. Thorne, {\it
  {Parton distributions from LHC, HERA, Tevatron and fixed target data: MSHT20
  PDFs}},  {\em Eur. Phys. J. C} {\bf 81} (2021) 341,
  [\href{http://arxiv.org/abs/2012.04684}{{\tt arXiv:2012.04684}}].

\bibitem{NNPDF:2021njg}
{\bf NNPDF} Collaboration, R.~D. Ball et~al., {\it {The path to proton
  structure at 1{\%} accuracy}},  {\em Eur. Phys. J. C} {\bf 82} (2022) 428,
  [\href{http://arxiv.org/abs/2109.02653}{{\tt arXiv:2109.02653}}].

\bibitem{PDF4LHCWorkingGroup:2022cjn}
{\bf PDF4LHC Working Group} Collaboration, R.~D. Ball et~al., {\it {The
  PDF4LHC21 combination of global PDF fits for the LHC Run III}},  {\em J.
  Phys. G} {\bf 49} (2022) 080501, [\href{http://arxiv.org/abs/2203.05506}{{\tt
  arXiv:2203.05506}}].

\bibitem{H1:2015ubc}
{\bf H1, ZEUS} Collaboration, H.~Abramowicz et~al., {\it {Combination of
  measurements of inclusive deep inelastic ${e^{\pm }p}$ scattering cross
  sections and QCD analysis of HERA data}},  {\em Eur. Phys. J. C} {\bf 75}
  (2015) 580, [\href{http://arxiv.org/abs/1506.06042}{{\tt arXiv:1506.06042}}].

\bibitem{Baur:1993zd}
U.~Baur, F.~Halzen, S.~Keller, M.~L. Mangano, and K.~Riesselmann, {\it {The
  Charm content of $W$ + 1 jet events as a probe of the strange quark
  distribution function}},  {\em Phys. Lett. B} {\bf 318} (1993) 544--548,
  [\href{http://arxiv.org/abs/hep-ph/9308370}{{\tt hep-ph/9308370}}].

\bibitem{ATLAS:2014jkm}
{\bf ATLAS} Collaboration, G.~Aad et~al., {\it {Measurement of the production
  of a $W$ boson in association with a charm quark in $pp$ collisions at
  $\sqrt{s} =$ 7 TeV with the ATLAS detector}},  {\em JHEP} {\bf 05} (2014)
  068, [\href{http://arxiv.org/abs/1402.6263}{{\tt arXiv:1402.6263}}].

\bibitem{CMS:2018dxg}
{\bf CMS} Collaboration, A.~M. Sirunyan et~al., {\it {Measurement of associated
  production of a W boson and a charm quark in proton-proton collisions at
  $\sqrt{s} =$ 13 TeV}},  {\em Eur. Phys. J. C} {\bf 79} (2019) 269,
  [\href{http://arxiv.org/abs/1811.10021}{{\tt arXiv:1811.10021}}].

\bibitem{CMS:2021oxn}
{\bf CMS} Collaboration, A.~Tumasyan et~al., {\it {Measurements of the
  associated production of a W boson and a charm quark in
  proton{\textendash}proton collisions at $\sqrt{s}=8\,\text {TeV} $}},  {\em
  Eur. Phys. J. C} {\bf 82} (2022) 1094,
  [\href{http://arxiv.org/abs/2112.00895}{{\tt arXiv:2112.00895}}].

\bibitem{ATLAS:2023ibp}
{\bf ATLAS} Collaboration, {ATLAS Collaboration}, {\it {Measurement of the
  production of a $W$ boson in association with a charmed hadron in $pp$
  collisions at $\sqrt{s} = 13\,\mathrm{TeV}$ with the ATLAS detector}},  {\em
  Phys. Rev. D} {\bf 108} (2023) 032012,
  [\href{http://arxiv.org/abs/2302.00336}{{\tt arXiv:2302.00336}}].

\bibitem{CMS:2023aim}
{\bf CMS} Collaboration, A.~Tumasyan et~al., {\it {Measurement of the
  production cross section for a W boson in association with a charm quark in
  proton{\textendash}proton collisions at $\sqrt{s} = 13\,\hbox {TeV}$}},  {\em
  Eur. Phys. J. C} {\bf 84} (2024) 27,
  [\href{http://arxiv.org/abs/2308.02285}{{\tt arXiv:2308.02285}}].

\bibitem{Catani:2004nc}
S.~Catani, D.~de~Florian, G.~Rodrigo, and W.~Vogelsang, {\it {Perturbative
  generation of a strange-quark asymmetry in the nucleon}},  {\em Phys. Rev.
  Lett.} {\bf 93} (2004) 152003,
  [\href{http://arxiv.org/abs/hep-ph/0404240}{{\tt hep-ph/0404240}}].

\bibitem{Stirling:2012vh}
W.~J. Stirling and E.~Vryonidou, {\it {Charm production in association with an
  electroweak gauge boson at the LHC}},  {\em Phys. Rev. Lett.} {\bf 109}
  (2012) 082002, [\href{http://arxiv.org/abs/1203.6781}{{\tt
  arXiv:1203.6781}}].

\bibitem{Czakon:2020coa}
M.~Czakon, A.~Mitov, M.~Pellen, and R.~Poncelet, {\it {NNLO QCD predictions for
  W+c-jet production at the LHC}},  {\em JHEP} {\bf 06} (2021) 100,
  [\href{http://arxiv.org/abs/2011.01011}{{\tt arXiv:2011.01011}}].

\bibitem{Czakon:2022khx}
M.~Czakon, A.~Mitov, M.~Pellen, and R.~Poncelet, {\it {A detailed investigation
  of W+c-jet at the LHC}},  {\em JHEP} {\bf 02} (2023) 241,
  [\href{http://arxiv.org/abs/2212.00467}{{\tt arXiv:2212.00467}}].

\bibitem{Alwall:2014hca}
J.~Alwall, et~al., {\it {The automated computation of tree-level and
  next-to-leading order differential cross sections, and their matching to
  parton shower simulations}},  {\em JHEP} {\bf 07} (2014) 079,
  [\href{http://arxiv.org/abs/1405.0301}{{\tt arXiv:1405.0301}}].

\bibitem{Bevilacqua:2021ovq}
G.~Bevilacqua, M.~V. Garzelli, A.~Kardos, and L.~Toth, {\it {W + charm
  production with massive c quarks in PowHel}},  {\em JHEP} {\bf 04} (2022)
  056, [\href{http://arxiv.org/abs/2106.11261}{{\tt arXiv:2106.11261}}].

\bibitem{Sherpa:2024mfk}
{\bf Sherpa} Collaboration, E.~Bothmann et~al., {\it {Event generation with
  Sherpa 3}},  {\em JHEP} {\bf 12} (2024) 156,
  [\href{http://arxiv.org/abs/2410.22148}{{\tt arXiv:2410.22148}}].

\bibitem{Bellm:2015jjp}
J.~Bellm et~al., {\it {Herwig 7.0/Herwig++ 3.0 release note}},  {\em Eur. Phys.
  J. C} {\bf 76} (2016) 196, [\href{http://arxiv.org/abs/1512.01178}{{\tt
  arXiv:1512.01178}}].

\bibitem{Banfi:2006hf}
A.~Banfi, G.~P. Salam, and G.~Zanderighi, {\it {Infrared safe definition of jet
  flavor}},  {\em Eur. Phys. J. C} {\bf 47} (2006) 113--124,
  [\href{http://arxiv.org/abs/hep-ph/0601139}{{\tt hep-ph/0601139}}].

\bibitem{Caletti:2022hnc}
S.~Caletti, A.~J. Larkoski, S.~Marzani, and D.~Reichelt, {\it {Practical jet
  flavour through NNLO}},  {\em Eur. Phys. J. C} {\bf 82} (2022) 632,
  [\href{http://arxiv.org/abs/2205.01109}{{\tt arXiv:2205.01109}}].

\bibitem{Caletti:2022glq}
S.~Caletti, A.~J. Larkoski, S.~Marzani, and D.~Reichelt, {\it {A fragmentation
  approach to jet flavor}},  {\em JHEP} {\bf 10} (2022) 158,
  [\href{http://arxiv.org/abs/2205.01117}{{\tt arXiv:2205.01117}}].

\bibitem{Czakon:2022wam}
M.~Czakon, A.~Mitov, and R.~Poncelet, {\it {Infrared-safe flavoured
  anti-k$_{T}$ jets}},  {\em JHEP} {\bf 04} (2023) 138,
  [\href{http://arxiv.org/abs/2205.11879}{{\tt arXiv:2205.11879}}].

\bibitem{Gauld:2022lem}
R.~Gauld, A.~Huss, and G.~Stagnitto, {\it {Flavor Identification of
  Reconstructed Hadronic Jets}},  {\em Phys. Rev. Lett.} {\bf 130} (2023)
  161901, [\href{http://arxiv.org/abs/2208.11138}{{\tt arXiv:2208.11138}}].
  [Erratum: Phys.Rev.Lett. 132, 159901 (2024)].

\bibitem{Caola:2023wpj}
F.~Caola, et~al., {\it {Flavored jets with exact anti-kt kinematics and tests
  of infrared and collinear safety}},  {\em Phys. Rev. D} {\bf 108} (2023)
  094010, [\href{http://arxiv.org/abs/2306.07314}{{\tt arXiv:2306.07314}}].

\bibitem{Behring:2025ilo}
A.~Behring et~al., {\it {Flavoured jet algorithms: a comparative study}},  {\em
  JHEP} {\bf 09} (2025) 149, [\href{http://arxiv.org/abs/2506.13449}{{\tt
  arXiv:2506.13449}}].

\bibitem{Gribov:1972ri}
V.~N. Gribov and L.~N. Lipatov, {\it {Deep inelastic e p scattering in
  perturbation theory}},  {\em Sov. J. Nucl. Phys.} {\bf 15} (1972) 438--450.

\bibitem{Altarelli:1977zs}
G.~Altarelli and G.~Parisi, {\it {Asymptotic Freedom in Parton Language}},
  {\em Nucl. Phys. B} {\bf 126} (1977) 298--318.

\bibitem{Dokshitzer:1977sg}
Y.~L. Dokshitzer, {\it {Calculation of the Structure Functions for Deep
  Inelastic Scattering and e+ e- Annihilation by Perturbation Theory in Quantum
  Chromodynamics.}},  {\em Sov. Phys. JETP} {\bf 46} (1977) 641--653.

\bibitem{Berman:1971xz}
S.~M. Berman, J.~D. Bjorken, and J.~B. Kogut, {\it {Inclusive Processes at High
  Transverse Momentum}},  {\em Phys. Rev. D} {\bf 4} (1971) 3388.

\bibitem{Rijken:1996vr}
P.~J. Rijken and W.~L. van Neerven, {\it {O (alpha-s**2) contributions to the
  longitudinal fragmentation function in e+ e- annihilation}},  {\em Phys.
  Lett. B} {\bf 386} (1996) 422--428,
  [\href{http://arxiv.org/abs/hep-ph/9604436}{{\tt hep-ph/9604436}}].

\bibitem{Rijken:1996ns}
P.~J. Rijken and W.~L. van Neerven, {\it {Higher order QCD corrections to the
  transverse and longitudinal fragmentation functions in electron - positron
  annihilation}},  {\em Nucl. Phys. B} {\bf 487} (1997) 233--282,
  [\href{http://arxiv.org/abs/hep-ph/9609377}{{\tt hep-ph/9609377}}].

\bibitem{Rijken:1996npa}
P.~J. Rijken and W.~L. van Neerven, {\it {O (alpha-s**2) contributions to the
  asymmetric fragmentation function in e+ e- annihilation}},  {\em Phys. Lett.
  B} {\bf 392} (1997) 207--215,
  [\href{http://arxiv.org/abs/hep-ph/9609379}{{\tt hep-ph/9609379}}].

\bibitem{Mitov:2006wy}
A.~Mitov and S.-O. Moch, {\it {QCD Corrections to Semi-Inclusive Hadron
  Production in Electron-Positron Annihilation at Two Loops}},  {\em Nucl.
  Phys. B} {\bf 751} (2006) 18--52,
  [\href{http://arxiv.org/abs/hep-ph/0604160}{{\tt hep-ph/0604160}}].

\bibitem{Blumlein:2006rr}
J.~Blumlein and V.~Ravindran, {\it {O (alpha**2(s)) Timelike Wilson
  Coefficients for Parton-Fragmentation Functions in Mellin Space}},  {\em
  Nucl. Phys. B} {\bf 749} (2006) 1--24,
  [\href{http://arxiv.org/abs/hep-ph/0604019}{{\tt hep-ph/0604019}}].

\bibitem{Goyal:2023zdi}
S.~Goyal, S.-O. Moch, V.~Pathak, N.~Rana, and V.~Ravindran, {\it
  {Next-to-Next-to-Leading Order QCD Corrections to Semi-Inclusive
  Deep-Inelastic Scattering}},  {\em Phys. Rev. Lett.} {\bf 132} (2024) 251902,
  [\href{http://arxiv.org/abs/2312.17711}{{\tt arXiv:2312.17711}}].

\bibitem{Bonino:2024qbh}
L.~Bonino, T.~Gehrmann, and G.~Stagnitto, {\it {Semi-Inclusive Deep-Inelastic
  Scattering at Next-to-Next-to-Leading Order in QCD}},  {\em Phys. Rev. Lett.}
  {\bf 132} (2024) 251901, [\href{http://arxiv.org/abs/2401.16281}{{\tt
  arXiv:2401.16281}}].

\bibitem{Bonino:2024wgg}
L.~Bonino, T.~Gehrmann, M.~L{\"o}chner, K.~Sch{\"o}nwald, and G.~Stagnitto,
  {\it {Polarized Semi-Inclusive Deep-Inelastic Scattering at
  Next-to-Next-to-Leading Order in QCD}},  {\em Phys. Rev. Lett.} {\bf 133}
  (2024) 211904, [\href{http://arxiv.org/abs/2404.08597}{{\tt
  arXiv:2404.08597}}].

\bibitem{Goyal:2024tmo}
S.~Goyal, et~al., {\it {Next-to-Next-to-Leading Order QCD Corrections to
  Polarized Semi-Inclusive Deep-Inelastic Scattering}},  {\em Phys. Rev. Lett.}
  {\bf 133} (2024) 211905, [\href{http://arxiv.org/abs/2404.09959}{{\tt
  arXiv:2404.09959}}].

\bibitem{Ahmed:2024owh}
T.~Ahmed, et~al., {\it {NNLO phase-space integrals for semi-inclusive
  deep-inelastic scattering}},  {\em Phys. Rev. D} {\bf 112} (2025) 014020,
  [\href{http://arxiv.org/abs/2412.16509}{{\tt arXiv:2412.16509}}].

\bibitem{Goyal:2024emo}
S.~Goyal, et~al., {\it {NNLO QCD corrections to unpolarized and polarized
  SIDIS}},  {\em Phys. Rev. D} {\bf 111} (2025) 094007,
  [\href{http://arxiv.org/abs/2412.19309}{{\tt arXiv:2412.19309}}].

\bibitem{Bonino:2025tnf}
L.~Bonino, T.~Gehrmann, M.~L{\"o}chner, K.~Sch{\"o}nwald, and G.~Stagnitto,
  {\it {Identified Hadron Production in Deeply Inelastic Neutrino-Nucleon
  Scattering}},  \href{http://arxiv.org/abs/2504.05376}{{\tt
  arXiv:2504.05376}}.

\bibitem{Bonino:2025qta}
L.~Bonino, T.~Gehrmann, M.~L{\"o}chner, K.~Sch{\"o}nwald, and G.~Stagnitto,
  {\it {Neutral and charged current semi-inclusive deep-inelastic scattering at
  NNLO QCD}},  {\em JHEP} {\bf 10} (2025) 016,
  [\href{http://arxiv.org/abs/2506.19926}{{\tt arXiv:2506.19926}}].

\bibitem{Goyal:2025bzf}
S.~Goyal, S.-O. Moch, V.~Pathak, N.~Rana, and V.~Ravindran, {\it {Soft and
  virtual corrections to semi-inclusive DIS up to four loops in QCD}},
  \href{http://arxiv.org/abs/2506.24078}{{\tt arXiv:2506.24078}}.

\bibitem{Bonino:2025bqa}
L.~Bonino, T.~Gehrmann, M.~L{\"o}chner, K.~Sch{\"o}nwald, and G.~Stagnitto,
  {\it {Polarized Neutral and Charged Current Semi-Inclusive Deep-Inelastic
  Scattering at NNLO in QCD}},  \href{http://arxiv.org/abs/2510.00100}{{\tt
  arXiv:2510.00100}}.

\bibitem{Goyal:2025qyu}
S.~Goyal, R.~N. Lee, S.-O. Moch, V.~Pathak, and V.~Ravindran, {\it {NNLO
  QCD$\otimes$QED corrections to unpolarized and polarized SIDIS}},
  \href{http://arxiv.org/abs/2510.18872}{{\tt arXiv:2510.18872}}.

\bibitem{Czakon:2021ohs}
M.~Czakon, T.~Generet, A.~Mitov, and R.~Poncelet, {\it {B-hadron production in
  NNLO QCD: application to LHC t$ \overline{t} $ events with leptonic decays}},
   {\em JHEP} {\bf 10} (2021) 216, [\href{http://arxiv.org/abs/2102.08267}{{\tt
  arXiv:2102.08267}}].

\bibitem{Czakon:2022pyz}
M.~Czakon, T.~Generet, A.~Mitov, and R.~Poncelet, {\it {NNLO B-fragmentation
  fits and their application to $ t\overline{t} $ production and decay at the
  LHC}},  {\em JHEP} {\bf 03} (2023) 251,
  [\href{http://arxiv.org/abs/2210.06078}{{\tt arXiv:2210.06078}}].

\bibitem{Bonino:2024adk}
L.~Bonino, T.~Gehrmann, M.~Marcoli, R.~Sch{\"u}rmann, and G.~Stagnitto, {\it
  {Antenna subtraction for processes with identified particles at hadron
  colliders}},  {\em JHEP} {\bf 08} (2024) 073,
  [\href{http://arxiv.org/abs/2406.09925}{{\tt arXiv:2406.09925}}].

\bibitem{Czakon:2024tjr}
M.~Czakon, T.~Generet, A.~Mitov, and R.~Poncelet, {\it {Open B-Hadron
  Production at Hadron Colliders in QCD at Next-to-Next-to-Leading-Order and
  Next-to-Next-to-Leading-Logarithmic Accuracy}},  {\em Phys. Rev. Lett.} {\bf
  135} (2025) 161903, [\href{http://arxiv.org/abs/2411.09684}{{\tt
  arXiv:2411.09684}}].

\bibitem{Czakon:2025yti}
M.~Czakon, T.~Generet, A.~Mitov, and R.~Poncelet, {\it {Identified Hadron
  Production at Hadron Colliders in Next-to-Next-to-Leading-Order QCD}},  {\em
  Phys. Rev. Lett.} {\bf 135} (2025) 17,
  [\href{http://arxiv.org/abs/2503.11489}{{\tt arXiv:2503.11489}}].

\bibitem{McGowan:2022nag}
J.~McGowan, T.~Cridge, L.~A. Harland-Lang, and R.~S. Thorne, {\it {Approximate
  N$^{3}$LO parton distribution functions with theoretical uncertainties:
  MSHT20aN$^3$LO PDFs}},  {\em Eur. Phys. J. C} {\bf 83} (2023) 185,
  [\href{http://arxiv.org/abs/2207.04739}{{\tt arXiv:2207.04739}}]. [Erratum:
  Eur.Phys.J.C 83, 302 (2023)].

\bibitem{NNPDF:2024nan}
{\bf NNPDF} Collaboration, R.~D. Ball et~al., {\it {The path to $\hbox
  {N}^3\hbox {LO}$ parton distributions}},  {\em Eur. Phys. J. C} {\bf 84}
  (2024) 659, [\href{http://arxiv.org/abs/2402.18635}{{\tt arXiv:2402.18635}}].

\bibitem{Czakon:2010td}
M.~Czakon, {\it {A novel subtraction scheme for double-real radiation at
  NNLO}},  {\em Phys. Lett. B} {\bf 693} (2010) 259--268,
  [\href{http://arxiv.org/abs/1005.0274}{{\tt arXiv:1005.0274}}].

\bibitem{Czakon:2014oma}
M.~Czakon and D.~Heymes, {\it {Four-dimensional formulation of the
  sector-improved residue subtraction scheme}},  {\em Nucl. Phys. B} {\bf 890}
  (2014) 152--227, [\href{http://arxiv.org/abs/1408.2500}{{\tt
  arXiv:1408.2500}}].

\bibitem{Czakon:2019tmo}
M.~Czakon, A.~van Hameren, A.~Mitov, and R.~Poncelet, {\it {Single-jet
  inclusive rates with exact color at $ \mathcal{O} $ ($ {\alpha}_s^4 $)}},
  {\em JHEP} {\bf 10} (2019) 262, [\href{http://arxiv.org/abs/1907.12911}{{\tt
  arXiv:1907.12911}}].

\bibitem{Bury:2015dla}
M.~Bury and A.~van Hameren, {\it {Numerical evaluation of multi-gluon
  amplitudes for High Energy Factorization}},  {\em Comput. Phys. Commun.} {\bf
  196} (2015) 592--598, [\href{http://arxiv.org/abs/1503.08612}{{\tt
  arXiv:1503.08612}}].

\bibitem{Cascioli:2011va}
F.~Cascioli, P.~Maierhofer, and S.~Pozzorini, {\it {Scattering Amplitudes with
  Open Loops}},  {\em Phys. Rev. Lett.} {\bf 108} (2012) 111601,
  [\href{http://arxiv.org/abs/1111.5206}{{\tt arXiv:1111.5206}}].

\bibitem{Buccioni:2017yxi}
F.~Buccioni, S.~Pozzorini, and M.~Zoller, {\it {On-the-fly reduction of open
  loops}},  {\em Eur. Phys. J. C} {\bf 78} (2018) 70,
  [\href{http://arxiv.org/abs/1710.11452}{{\tt arXiv:1710.11452}}].

\bibitem{Buccioni:2019sur}
F.~Buccioni, et~al., {\it {OpenLoops 2}},  {\em Eur. Phys. J. C} {\bf 79}
  (2019) 866, [\href{http://arxiv.org/abs/1907.13071}{{\tt arXiv:1907.13071}}].

\bibitem{Gehrmann:2011ab}
T.~Gehrmann and L.~Tancredi, {\it {Two-loop QCD helicity amplitudes for $q\bar
  q \to W^\pm \gamma$ and $q\bar q \to Z^0 \gamma$}},  {\em JHEP} {\bf 02}
  (2012) 004, [\href{http://arxiv.org/abs/1112.1531}{{\tt arXiv:1112.1531}}].

\bibitem{Bauer:2000cp}
C.~W. Bauer, A.~Frink, and R.~Kreckel, {\it {Introduction to the GiNaC
  framework for symbolic computation within the C++ programming language}},
  {\em J. Symb. Comput.} {\bf 33} (2002) 1--12,
  [\href{http://arxiv.org/abs/cs/0004015}{{\tt cs/0004015}}].

\bibitem{Vollinga:2004sn}
J.~Vollinga and S.~Weinzierl, {\it {Numerical evaluation of multiple
  polylogarithms}},  {\em Comput. Phys. Commun.} {\bf 167} (2005) 177,
  [\href{http://arxiv.org/abs/hep-ph/0410259}{{\tt hep-ph/0410259}}].

\bibitem{ParticleDataGroup:2020ssz}
{\bf Particle Data Group} Collaboration, P.~A. Zyla et~al., {\it {Review of
  Particle Physics}},  {\em PTEP} {\bf 2020} (2020) 083C01.

\bibitem{Buckley:2014ana}
A.~Buckley, et~al., {\it {LHAPDF6: parton density access in the LHC precision
  era}},  {\em Eur. Phys. J. C} {\bf 75} (2015) 132,
  [\href{http://arxiv.org/abs/1412.7420}{{\tt arXiv:1412.7420}}].

\bibitem{ATLAS:2021vod}
{\bf ATLAS} Collaboration, {ATLAS Collaboration}, {\it {Determination of the
  parton distribution functions of the proton using diverse ATLAS data from
  $pp$ collisions at $\sqrt{s} = 7$, 8 and 13~TeV}},  {\em Eur. Phys. J. C}
  {\bf 82} (2022) 438, [\href{http://arxiv.org/abs/2112.11266}{{\tt
  arXiv:2112.11266}}].

\bibitem{Soleymaninia:2017xhc}
M.~Soleymaninia, H.~Khanpour, and S.~M. Moosavi~Nejad, {\it {First
  determination of $D^{*+}$-meson fragmentation functions and their
  uncertainties at next-to-next-to-leading order}},  {\em Phys. Rev. D} {\bf
  97} (2018) 074014, [\href{http://arxiv.org/abs/1711.11344}{{\tt
  arXiv:1711.11344}}].

\bibitem{Salajegheh:2019nea}
M.~Salajegheh, S.~M. Moosavi~Nejad, M.~Soleymaninia, H.~Khanpour, and
  S.~Atashbar~Tehrani, {\it {NNLO charmed-meson fragmentation functions and
  their uncertainties in the presence of meson mass corrections}},  {\em Eur.
  Phys. J. C} {\bf 79} (2019) 999, [\href{http://arxiv.org/abs/1904.09832}{{\tt
  arXiv:1904.09832}}].

\bibitem{Anderle:2017cgl}
D.~P. Anderle, T.~Kaufmann, M.~Stratmann, F.~Ringer, and I.~Vitev, {\it {Using
  hadron-in-jet data in a global analysis of $D^{*}$ fragmentation functions}},
   {\em Phys. Rev. D} {\bf 96} (2017) 034028,
  [\href{http://arxiv.org/abs/1706.09857}{{\tt arXiv:1706.09857}}].

\bibitem{Bertone:2013vaa}
V.~Bertone, S.~Carrazza, and J.~Rojo, {\it {APFEL: A PDF Evolution Library with
  QED corrections}},  {\em Comput. Phys. Commun.} {\bf 185} (2014) 1647--1668,
  [\href{http://arxiv.org/abs/1310.1394}{{\tt arXiv:1310.1394}}].

\bibitem{Mele:1990yq}
B.~Mele and P.~Nason, {\it {Next-to-leading QCD calculation of the heavy quark
  fragmentation function}},  {\em Phys. Lett. B} {\bf 245} (1990) 635--639.

\bibitem{Mele:1990cw}
B.~Mele and P.~Nason, {\it {The Fragmentation function for heavy quarks in
  QCD}},  {\em Nucl. Phys. B} {\bf 361} (1991) 626--644. [Erratum: Nucl.Phys.B
  921, 841--842 (2017)].

\bibitem{Bonino:2023icn}
L.~Bonino, M.~Cacciari, and G.~Stagnitto, {\it {Heavy quark fragmentation in
  e$^{+}$e$^{-}$ collisions to NNLO+NNLL accuracy in perturbative QCD}},  {\em
  JHEP} {\bf 06} (2024) 040, [\href{http://arxiv.org/abs/2312.12519}{{\tt
  arXiv:2312.12519}}].

\bibitem{CMS:2013wql}
{\bf CMS} Collaboration, {CMS Collaboration}, {\it {Measurement of Associated W
  + Charm Production in pp Collisions at $\sqrt{s}$ = 7 TeV}},  {\em JHEP} {\bf
  02} (2014) 013, [\href{http://arxiv.org/abs/1310.1138}{{\tt
  arXiv:1310.1138}}].

\end{thebibliography}\endgroup

\end{document}